\documentclass[final]{raa}            
\usepackage{natbib}
\usepackage{threeparttable}
\bibpunct{(}{)}{;}{a}{}{,}
\usepackage[pagebackref=true]{hyperref}
\hypersetup{pdftitle = The title of my PDF, pdfauthor = My name, pdfsubject= The subject, pdfkeywords = keyword1 keyword2 keyword3} 
\hypersetup{colorlinks = true, linkcolor = green, anchorcolor = red, citecolor = blue, filecolor = red, pagecolor = red, urlcolor = red}

\usepackage{graphicx,times}             
\usepackage{amsmath}
\usepackage{subfig}
\begin{document}

   \title{ Assessing the Performance of Molecular Gas Clump Identification Algorithms
\footnotetext{\small Supported by the National Key R\&D Program of China (NO. 2017YFA0402701).} 
}

 \volnopage{ {\bf 2012} Vol.\ {\bf X} No. {\bf XX}, 000--000}
   \setcounter{page}{1}

   \author{Chong Li\inst{1,2}, Hong-Chi Wang\inst{1}, Yuan-Wei Wu\inst{3}, Yue-Hui Ma
      \inst{1,2}, Liang-Hao Lin\inst{1,4}
   }

   \institute{ Purple Mountain Observatory and Key Laboratory of Radio Astronomy, Chinese Academy of Sciences, 10 Yuanhua Road, Nanjing 210033, China; {\it chongli@pmo.ac.cn}\\
        \and
             University of Chinese Academy of Sciences, 19A Yuquan Road, Shijingshan District, Beijing 100049, China\\
                     \and
             National Time Service Center, Key Laboratory of Precise Positioning and Timing Technology, Chinese Academy of Sciences, Xi$\arcmin$an 710600, China\\
             \and
			School of Astronomy and Space Science, University of Science and Technology of China,
			96 Jinzhai Road, Hefei 230026, China\\
\vs \no
   {\small Submitted 2019 April 18}
}

\abstract{The detection of clumps(cores) in molecular clouds is an important issue in sub-millimetre astronomy. However, the completeness of the identification and the accuracy of the returned parameters of the automated clump identification algorithms are still not clear by now. In this work, we test the performance and bias of the GaussClumps, ClumpFind, Fellwalker, Reinhold, and Dendrograms algorithms in identifying simulated clumps. By designing the simulated clumps with various sizes, peak brightness, and crowdedness, we investigate the characteristics of the algorithms and their performance. In the aspect of detection completeness, Fellwalker, Dendrograms, and Gaussclumps are the first, second, and third best algorithms, respectively. The numbers of correct identifications of the six algorithms gradually increase as the size and SNR of the simulated clumps increase and they decrease as the crowdedness increases. In the aspect of the accuracy of retrieved parameters, Fellwalker and Dendrograms exhibit better performance than the other algorithms. The average deviations in clump parameters for all algorithms gradually increase as the size and SNR of clumps increase. Most of the algorithms except Fellwalker exhibit significant deviation in extracting the total flux of clumps. Taken altogether, Fellwalker, Gaussclumps, and Dendrograms exhibit the best performance in detection completeness and extracting parameters. The deviation in virial parameter for the six algorithms is relatively low. When applying the six algorithms to the clump identification for the Rosette molecular cloud, ClumpFind1994, ClumpFind2006, Gaussclumps, Fellwalker, and Reinhold exhibit performance that is consistent  with the results from the simulated test.
}

   \authorrunning{Chong Li et al. }            
   \titlerunning{Comparison of the Clump Identification Algorithms}  
   \maketitle

%
\section{Introduction}           
\label{intro}

It has been widely known that giant molecular clouds (GMCs) have complex and hierarchical structures that can be divided into substructures of clouds, clumps, and cores \citep{1999ASIC..540....3B}. The clumps and cores could be gravitationally unstable \citep{1987ApJ...319..730S, 2001ApJ...551..852H} and evolve into protostars. A particular issue in sub-millimeter astronomy is the identification of clumps and cores. The traditional clumps(cores) identification method is to find compact and bright sources in observational datasets by eyes. In this case, subjective biases are evident as each person could perceive the data differently and thus identify different clumps and extract different parameters. As the datasets become larger or the clumps are more crowded, the traditional method is more inefficient or incompetent in the detection of clumps(cores).

Several common algorithms have been used to identify clumpy structures in molecular clouds, such as GaussClumps, ClumpFind, FellWalker, Reinhold, and Dendrograms. Except for Dendrograms, these algorithms are included within CUPID \citep{2007ASPC..376..425B} \footnote{\url{http://starlink.eao.hawaii.edu/starlink/CUPID}}. GaussClumps is the oldest algorithms in automated clump identification \citep{1990ApJ...356..513S}. It was first applied in the M17 molecular cloud and then was frequently performed in other molecular clouds \citep{1998A&A...338..262S,2009MNRAS.395.1805D,2009MNRAS.395.1021L}. The GaussClumps algorithm fits the 3D molecular line data with Gaussian ellipsoids (or ellipses for 2D column density maps) around the local maxima. The resulting ellipsoids(or ellipses) are recognized as clumps in the observational data. This process is repeated until the termination criteria are met. The output clumps may overlap in the GaussClumps algorithm. For this reason, each input pixel is not simply assigned to a single clump (as what is done in algorithms such as FellWalker or ClumpFind), and the total flux in the fitted Gaussians may exceed the real flux in the input data. The GaussClumps algorithm can only fit a strict elliptic shape and it does not allocate flux to a clump at large distance from the peak.

ClumpFind is the most widely used algorithm for molecular gas clump identification. \citet{1994ApJ...428..693W} developed this algorithm and applied it to detect the compact structures in the Rosette molecular cloud. In brief, this algorithm locates the peak position by determining the highest value in the array. Then the process descends down from the peak pixel with a certain interval (ClumpFind.DeltaT). If no new independent maximum is found within an intensity interval, the process continues into lower intensity intervals until a new local maximum is found or it reaches the specified minimum contour level (ClumpFind.TLow). Clumps with brightness below this level are ignored as they are assumed to be noise. When an area contains multiple clumps then the pixels in that area are divided between the clumps, with the association of each pixel being given to the closest clump. The association is determined by the distance of the pixel to the boundary of an assigned clump according to a friends-of-friends algorithm. The ClumpFind algorithm was re-written with some minor adjustments in 2006, so it has the ClumpFind1994 and ClumpFind2006 versions (set by ClumpFind.IDLAlg).

The ClumpFind algorithm is found to be sensitive to the input parameters ClumpFind.DeltaT and ClumpFind.TLow \citep{2009A&A...497..399K,2009ApJS..182..131R,2009ApJ...699L.134P}. A large ClumpFind.DeltaT parameter tends to find the large and bright structures but miss the clumps with low brightness. If a small ClumpFind.DeltaT value is provided, increased false clumps are identified due to the noise spike. \citet{2006ApJ...638..293E} performed investigations into the detection completeness of the ClumpFind algorithm. They found that ClumpFind tends to interpolate clumps and break the bright source into multiple clumps.

The Fellwalker algorithm was developed specifically for CUPID to address some of the problems associated with ClumpFind. It was developed and fully described in \citet{2015A&C....10...22B}. Unlike other algorithms, this algorithm firstly defines a minimum level (FellWalker.Noise) to ignore the influence of the noise spike. Then the process ascends the steepest route until reaching a peak, which provides the certain way of reaching the peak along the greatest ascending gradient. Sometimes this process may be affected by noise spikes and thus FellWalker checks the extended surrounding (FellWalker.MaxJump) pixels to see if there is a pixel in the surrounding with greater value. For any pixel on the minimum level, a path from this pixel to the nearby maximum value based on this ascending method can be found. All routes that meet at the same maximum point are classified as a clump. This process is analogous to a fell-walker ascending a hill by following the steepest ascent line as its name suggests. 

The Reinhold algorithm was developed by Kim Reinhold and included within CUPID. This algorithm converts the original two or three-dimensional data arrays into one-dimensional arrays. Then it identifies the highest value in all one-dimensional arrays. If the peak value is below the defined minimum then the algorithm decides that there is no real peak in that array. If the peak value is above the minimum then the program goes from this peak in both directions along the array until it reaches a pixel that fulfills the criteria for being an edge pixel. The data arrays are re-combined into the original two or three-dimensional arrays with the clump edges now determined, which produces a number of ring or shell-like structures which outline the clumps. Basically, the algorithm looks for edges of clumps, the clumps determined are therefore more susceptible to noise and need to be cleaned up.

The Dendrograms algorithm was first demonstrated in the structural analysis of molecular clouds by \citet{2008ApJ...679.1338R}. This algorithm presents an analytic technique aimed to characterize the hierarchical structure in molecular gas and relate it to the star formation process. Its principal advantage is using standard molecular line analysis techniques to characterize the branches in a dendrogram and simultaneously provide the measurement of various properties for structures in a large range of physical scales. The smallest structure which is described as leaves in Dendrograms can be recognized as clumps. In addition, Dendrograms is a reduction of the structure in a data set to its essential features. Three parameters (min\_value, min\_delta, and min\_npix) would limit the output results of the clump identification. It is the newest algorithm compared with the above other algorithms for structure identification but has been performed more than one hundred times \citep{2009Natur.457...63G,2019MNRAS.483.5135W}.

So far, many automated algorithms have been widely used for clump identification, and the principles are different. Different algorithms could make bias results in both clump identification and extraction of parameters such as size, line width, temperature, and mass \citep{2004PASA...21..290S,2010MNRAS.402..603C,2010MsT..........1W}. However, it is still not clear which algorithm has the best performance in the aspects of completeness, false detection probability, and accuracy of physical parameters of the clump identification. Simulated test is needed before applying these algorithms in observational data. In this work, we mainly focus on testing the completeness and accuracy of the physical parameters of the clumps identified using the above six algorithms (including two versions of ClumpFind) and present comparisons between them. The method is described in Section \ref{method} and the results are presented in Section \ref{result}. We discuss the bias of the algorithms in estimation of the virial parameter and the performances of algorithms in identifying clumps in the Rosette molecular cloud in Section \ref{discussion} and make a summary in Section \ref{summary}. 



\section{Method}           
\label{method}


Mass-size relations describe the relationship between the mass and spatial scale of clumps. The mass contained within radius $r$ is usually described with a power law $m(r) \sim r^{-k}$ \citep{1981MNRAS.194..809L,2007ARA&A..45..565M,2010ApJ...712.1137K,2010ApJ...716..433K}. This relation can be explained by a power law density profiles of the molecular clumps: $\rho(r) \sim r^{-p}$ \citep{2011MNRAS.416..783P}. Previous studies found that $1.5 \leq p \leq 2$ \citep{1993A&A...278..238H,2000A&A...357..637H,2002ApJ...566..945B,2002A&A...389..603F,2002ApJS..143..469M}. However, it has been found that a single power law density profile cannot fit the emission from starless cores and that an inner flattening portion is always needed to reproduce the observational data \citep{1994MNRAS.268..276W,1996A&A...314..625A,2000A&A...361..555B,1999ApJ...515..265A}. Considering the power law behavior for large $r$ and the central flattening at small $r$, \citet{2002ApJ...569..815T} adopted the following analytic density profile for molecular clumps,

\begin{equation}
\label{eqa1}
\rho(r) \sim \frac{1}{1+(r/r_0)^p}
\end{equation}

Here, $r_0$ is the radius of the flat region ($2r_0$ is the FWHM), $r =(s^2 + z^2)^{1/2}$, $s$ is the projected distance from the clump center and $z$ is the length along the line of sight. In this case, the column density distribution of the clump obeys

\begin{equation}
\label{eqa3}
N(s) \sim \int \frac{1}{1+(s/r_0)^p}\, dz
\end{equation}

\citet{2010ApJ...716..433K} found that if the  index of density profile is $p$, the  index of column density profile can be approximated as $p-1$ when $s>>r_0$. Then the radial column density profile follows

\begin{equation}
\label{eqa4}
N(s) \sim \frac{1}{1+(s/r_0)^{p-1}}
\end{equation}

\citet{2018A&A...614A..83J} have studied the column density structures of the Galactic Cold Cores. They found that the radial column density profiles of these cores follow power law distributions with the indexes of about 1. Therefor, we adopt a column density profile of Equation \ref{eqa4} with $p-1=1$ for clumps in this study.

The third dimension of observational data stands for the velocity. An optically thin spectrum of a clump probes the velocity distribution of the molecules. Indeed, the velocity profile of an optically thin molecular line emission in observations generally follows the Gaussian distribution. Therefor, the brightness distribution over the voxels of our simulated clumps is chosen to obey the form

\begin{equation}
\label{eqa2}
T(s,v) = N(s) \times \frac{1}{\sqrt{2\pi}\sigma}exp(\frac{-(v-v_0)^2}{2\sigma^2})= \frac{N_0}{\sqrt{2\pi}\sigma} \times \frac{1}{1+(s/r_0)^{1.0}} \times exp(\frac{-(v-v_0)^2}{2\sigma^2})
\end{equation}

where $\frac{N_0}{\sqrt{2\pi}\sigma}$ represents the peak brightness at the center of the clump.

We created three-dimensional arrays that contain clumps and background noise. The positions of the clumps are designed to distribute randomly. To avoid that the clumps are located at the edges of the arrays, the centers of the simulated clumps are distributed at least 15 pixels from the edges. The brightness profiles of the clumps are of the form of Equation \ref{eqa2} and we assume that $\sigma=r_0$ (in the unit of voxel). Thus, the FWHM(v) in the velocity dimension is equal to $\sqrt{8 \rm ln 2}r_0$, which is similar to the radial size of the clump ( FWHM(s) = $2r_0$ ). In the following analysis, the input size of the simulated clump is represented by the spatial FWHM size (FWHM(s) = $2r_0$). When we perform the identification of the compact sources in observational data, the FWHM, $\Delta$V, and the peak value of the clumps are naturally limited to the instrument resolution and the sensitivity of the observation. In our test, we set the FWHM and $\Delta$V to be 2 pixels and 2 channels, respectively, for all the six algorithms. The minimum peak value parameter is set to be 5 times the one sigma noise level and the number of voxels of an output clump is required to be above 16. We use other default input parameters in CUPID for all the six algorithms, so that we can determine the advantages and disadvantages of different algorithms. If the clumps identified by the algorithm are too far from the real situation, even if the results can be improved by adjusting the parameters later, it is clear that the algorithm is too sensitive to the parameters. Considering that it is a hard challenge for automated algorithms to identify clumps that are small, weak, or crowded, we created clumps with these characteristics in the test. Then the above six algorithms (GaussClumps, ClumpFind1994, ClumpFind2006, Fellwalker, Reinhold, and Dendrograms) are applied to identify the clumps. The configuration parameters of each algorithm are displayed in Appendix \ref{appendix} and their performance are presented in Section \ref{test1}-\ref{test3}. 

\section{Results}           
\label{result}

\subsection{Test 1: Performance of the Algorithms in Identifying Clumps of Different Sizes}
\label{test1}

\begin{figure}[h]
  \centering
\includegraphics[width=0.3\textwidth]{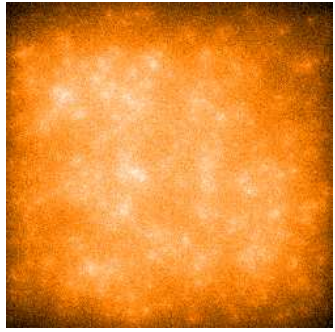}
    \caption{Simulated clumps with clump sizes (FWHM) distributing randomly from 1 to 11 pixels.}
  \label{fig:size}
\end{figure}

In order to investigate the performance of different algorithms in identifying clumps of different sizes, we generated 1000 clumps with clump sizes (FWHM = $2r_0$) distributing randomly from 1 to 11 pixels. The data are designed to be 1000$\times$1000$\times$1000 array (Figure \ref{fig:size}). In this case, very few clumps would overlap. The above six algorithms are applied to identify the clumps in the simulated data so that the performance of each algorithm can be unbiasedly estimated.

In test 1 we focus on the performance of the algorithms to detect different sizes of clumps. To reduce the influence of peak brightness, the  peak brightness of the clumps is fixed to be 10 times the one sigma noise value. We present the completeness and accuracy of the parameters of the six algorithms in Section \ref{completeness1} and \ref{accuracy1}.

\subsubsection{Completeness of the Algorithms}
\label{completeness1}

\begin{figure} [!htb]
\centering
\subfloat[]{\label{fig:right_size} 
\includegraphics[width=0.4\textwidth]{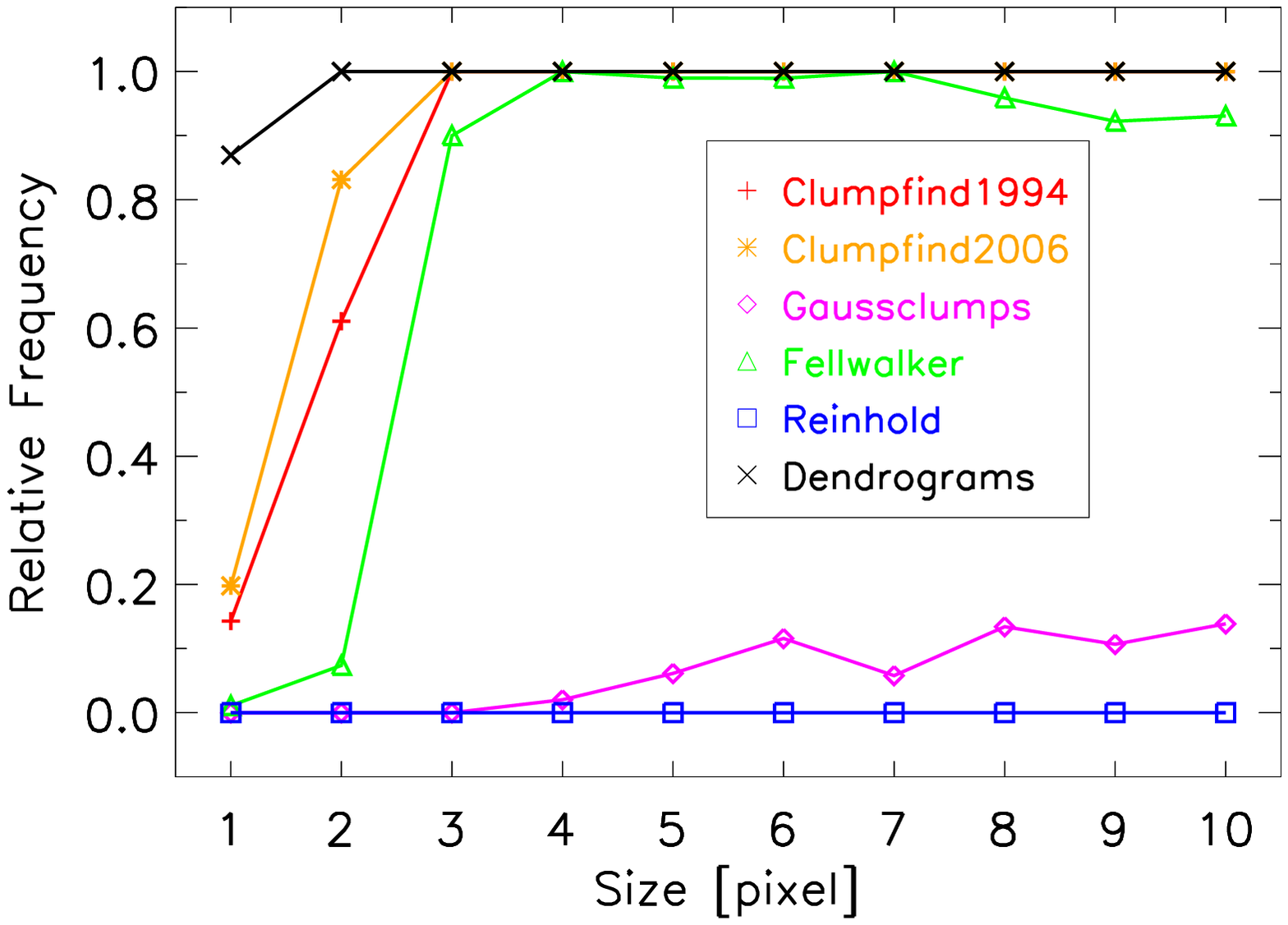}}
\subfloat[]{\label{fig:repeat_size}
\includegraphics[width=0.4\textwidth]{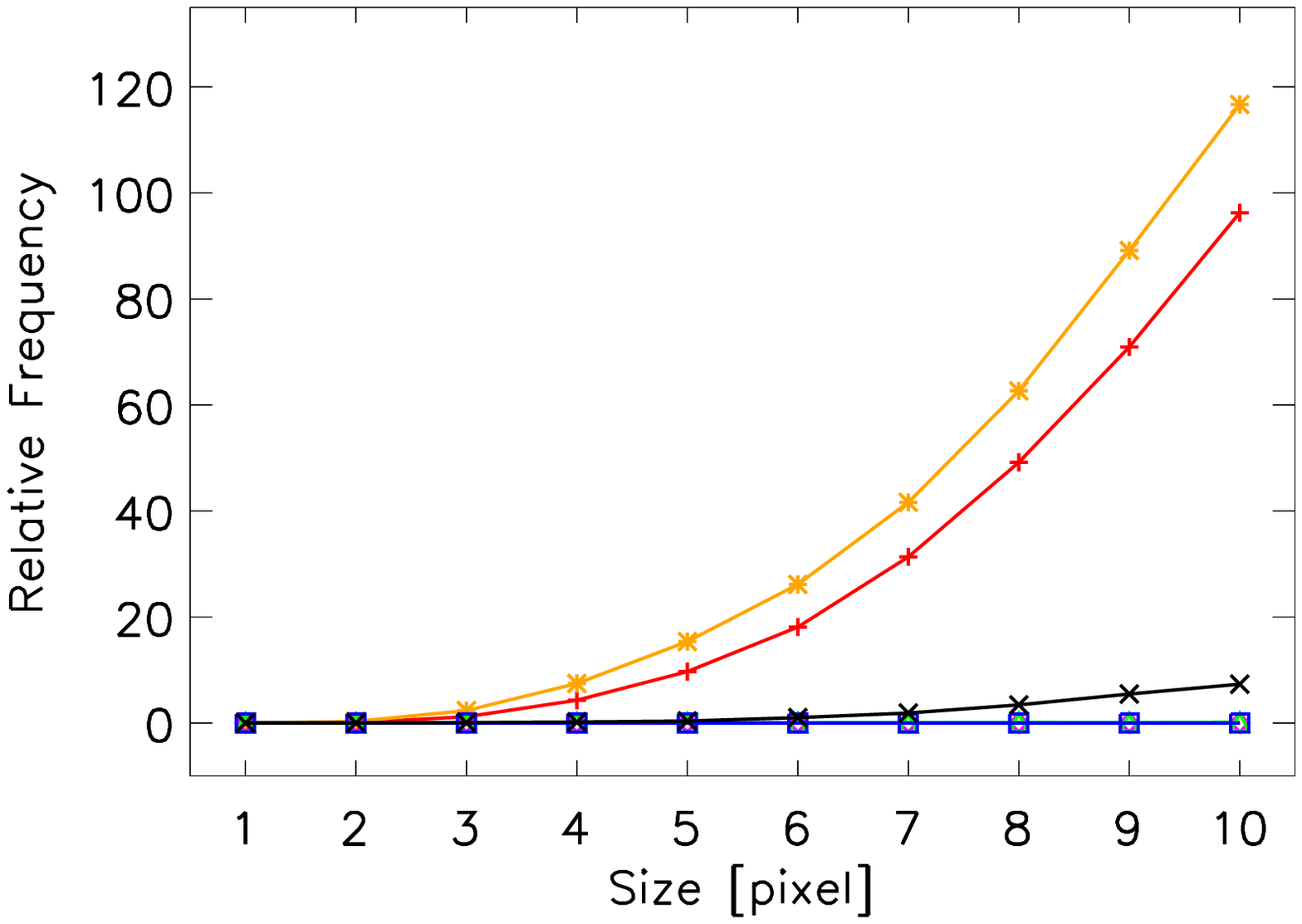}}\\
\subfloat[]{\label{fig:false_size}
\includegraphics[width=0.4\textwidth]{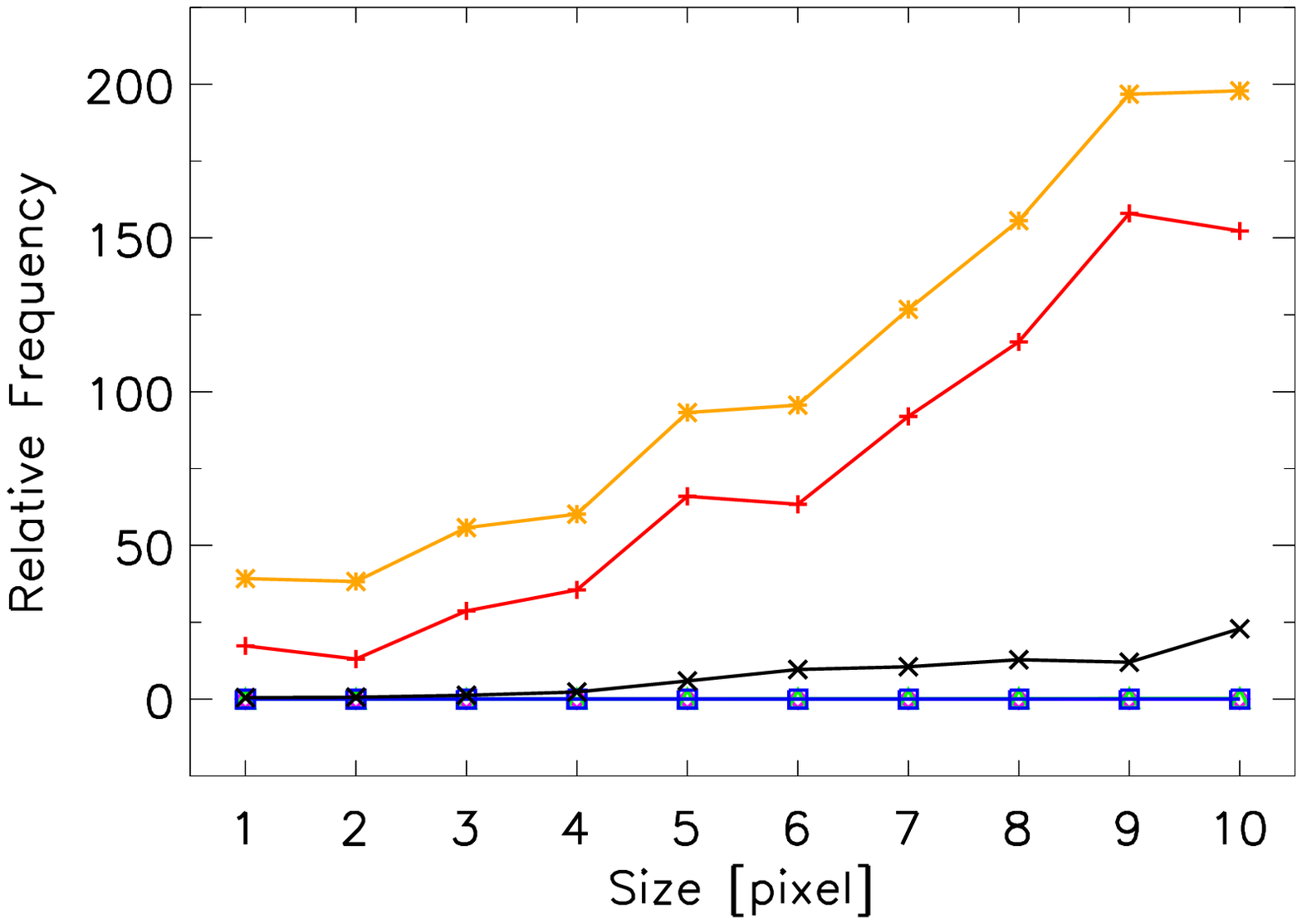}}
\subfloat[]{\label{fig:mark_size}
\includegraphics[width=0.4\textwidth]{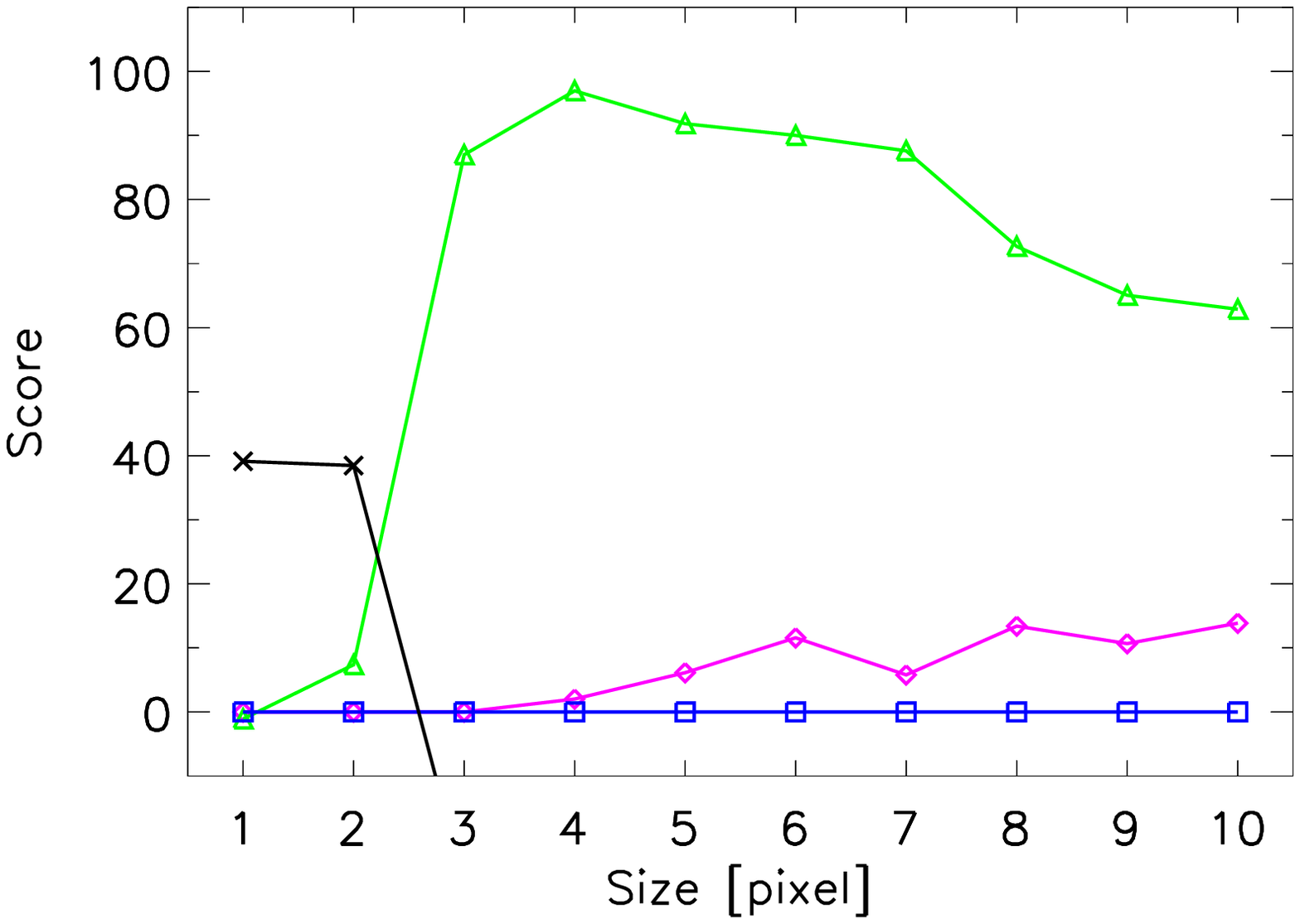}}
\caption{Frequency distributions of the correct identifications (a), repeated identifications (b), erroneous identifications (c), and score points (d) of the six algorithms in identifying clumps with different FWHM sizes. ClumpFind1994, ClumpFind2006, and Dendrograms score points lower than the range to display in panel (d).}
\label{fig:completeness_size}
\end{figure}

When estimating the advantages and disadvantages of an algorithm, the primary criterion is the completeness of the output clumps. Here the term "completeness" means a high percentage of correctly identified clumps, and at the meantime a low percentage of false or repeated identification. When the spatial scale of an output clump is smaller than the simulated clump along all the three axes, then we mark this clump as a correctly identified clump. Figure \ref{fig:right_size} shows the numbers of correct identifications for different algorithms when identifying clumps of different sizes. It can be seen that when the sizes of clumps are smaller than 2 pixels, most of the algorithms perform poorly. As clumps become larger, the numbers of correct identifications of the algorithms gradually increase. ClumpFind1994, ClumpFind2006, Fellwalker, and Dendrograms perform well when the sizes of clumps are larger than 3 pixels, where the correct rate can reach more than 90$\%$. Gaussclumps exhibits lower correct rate compared to its performance presented in Section \ref{test2}. One possible reason is that the brightness profile of simulated clumps obeys power law in the spatial dimensions.When the number of consecutive failures of fitting succeeds the designated one, which is appointed by parameter GaussClumps.MaxSkip, the iterative fitting process of Gaussclumps is terminated.

When a simulated clump is identified more than once, the clump closest to the simulated position is marked as a matching output and the other clumps are recorded as repeated results. Figure \ref{fig:repeat_size} shows the number of repeated identification for different algorithms. The percentages of repeated results for all algorithms are lower than 10$\%$ when the sizes of the input clumps are  smaller than 2 pixels. Figure \ref{fig:false_size} shows the number of erroneous identifications for the six algorithms. Fellwalker and Gaussclumps almost never erroneously identify the fluctuation of noise to be a clump. As the clumps become more extended, the numbers of repeated and erroneous identifications of the Dendrograms, ClumpFind1994, and ClumpFind2006 gradually increase.

In order to estimate the comprehensive performance of each algorithm, we establish a simple scoring mechanism. The algorithm scores 1 point when it correctly identifies a simulated clump, scores -1 when it outputs a false result, and score -0.5 when it finds duplicate clumps. We show the score for each algorithm in Figure \ref{fig:mark_size}. As shown in the scoring results, Fellwalker exhibits the best performance compared with the other algorithms. ClumpFind1994, ClumpFind2006, and Dendrograms receive low marks because they conduct a lot of repeated and erroneous identifications.

\subsubsection{Accuracy of Retrieved Parameters}
\label{accuracy1}

When an algorithm is used to automatically search for the clumps, the output parameters of the clumps will be used to calculated the physical parameters, so accurately reproducing the clump parameters is an important aspect of an algorithm. In order to compare the comprehensive performance of each algorithm in the aspect of accuracy of retrieved parameters, we calculated the average deviation of the position (E($|\Delta$X$|$)), size (E($\Delta$S)), velocity dispersion (E($\Delta$V)), peak brightness (E($\Delta$I)), and total flux (E($\Delta$flux)) of the output results of each algorithm. The E($|\Delta$X$|$), E($\Delta$S), E($\Delta$V), and E($\Delta$I) are obtained through subtracting the output parameters from the input ones and then averaging over the correctly identified clumps. The E($\Delta$flux) is the ratio between the deviation of the total flux and the input flux sum averaged over the correctly identified clumps.

\begin{figure} [!htb]
\centering
\subfloat[]{\label{fig:} 
\includegraphics[width=0.4\textwidth]{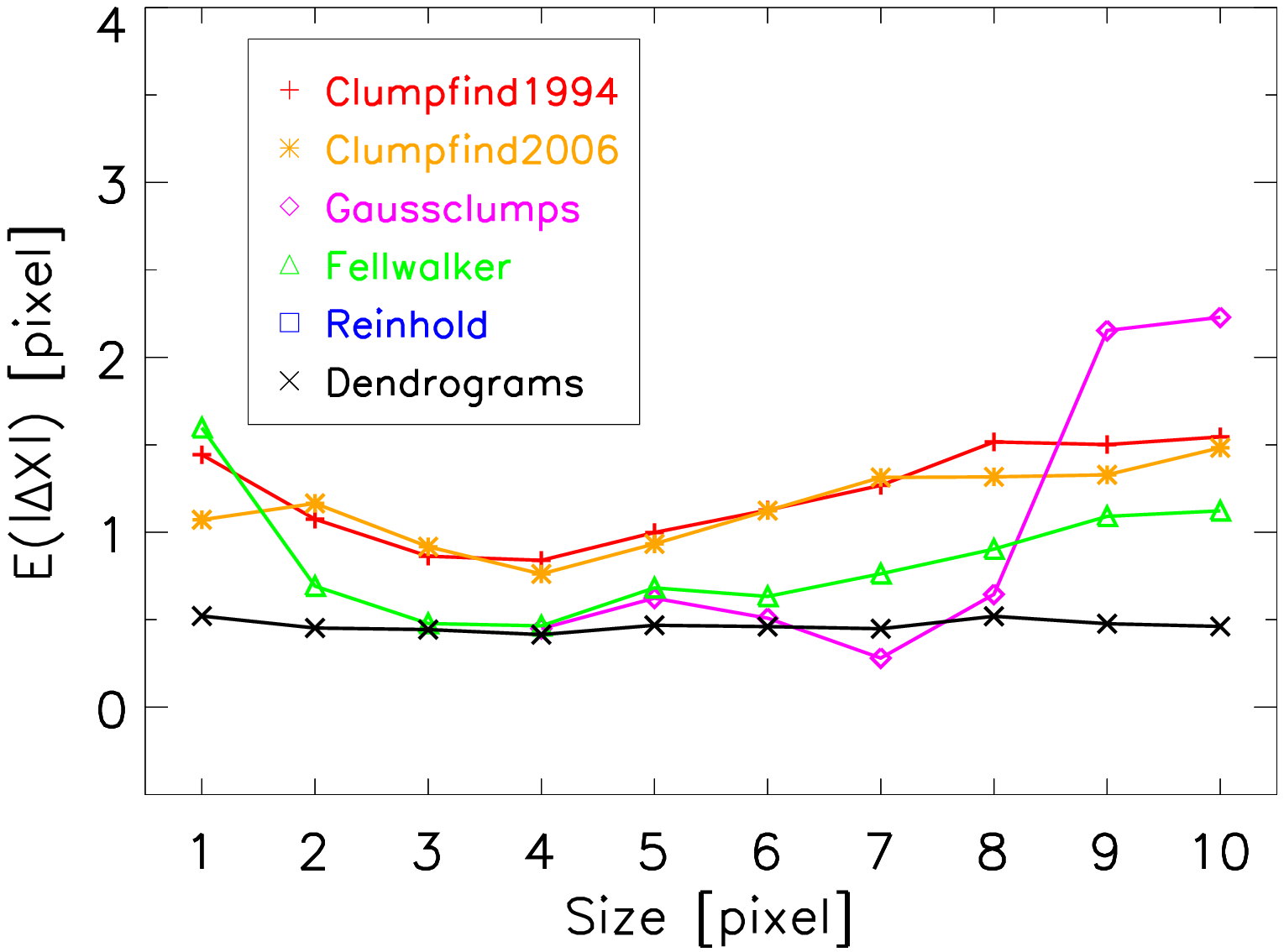}}
\subfloat[]{\label{fig:E_s_size}
\includegraphics[width=0.4\textwidth]{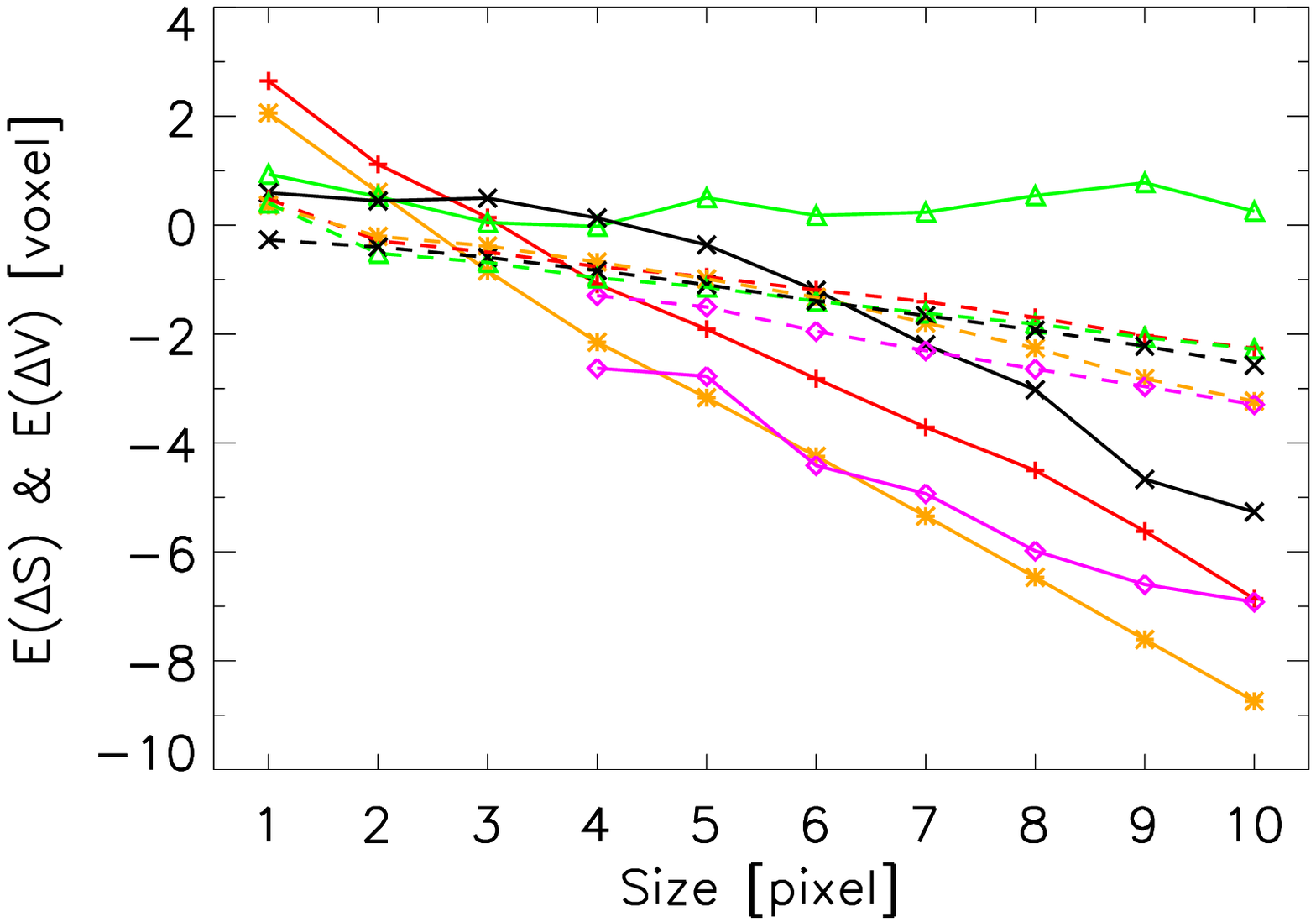}}\\
\subfloat[]{\label{fig:}
\includegraphics[width=0.4\textwidth]{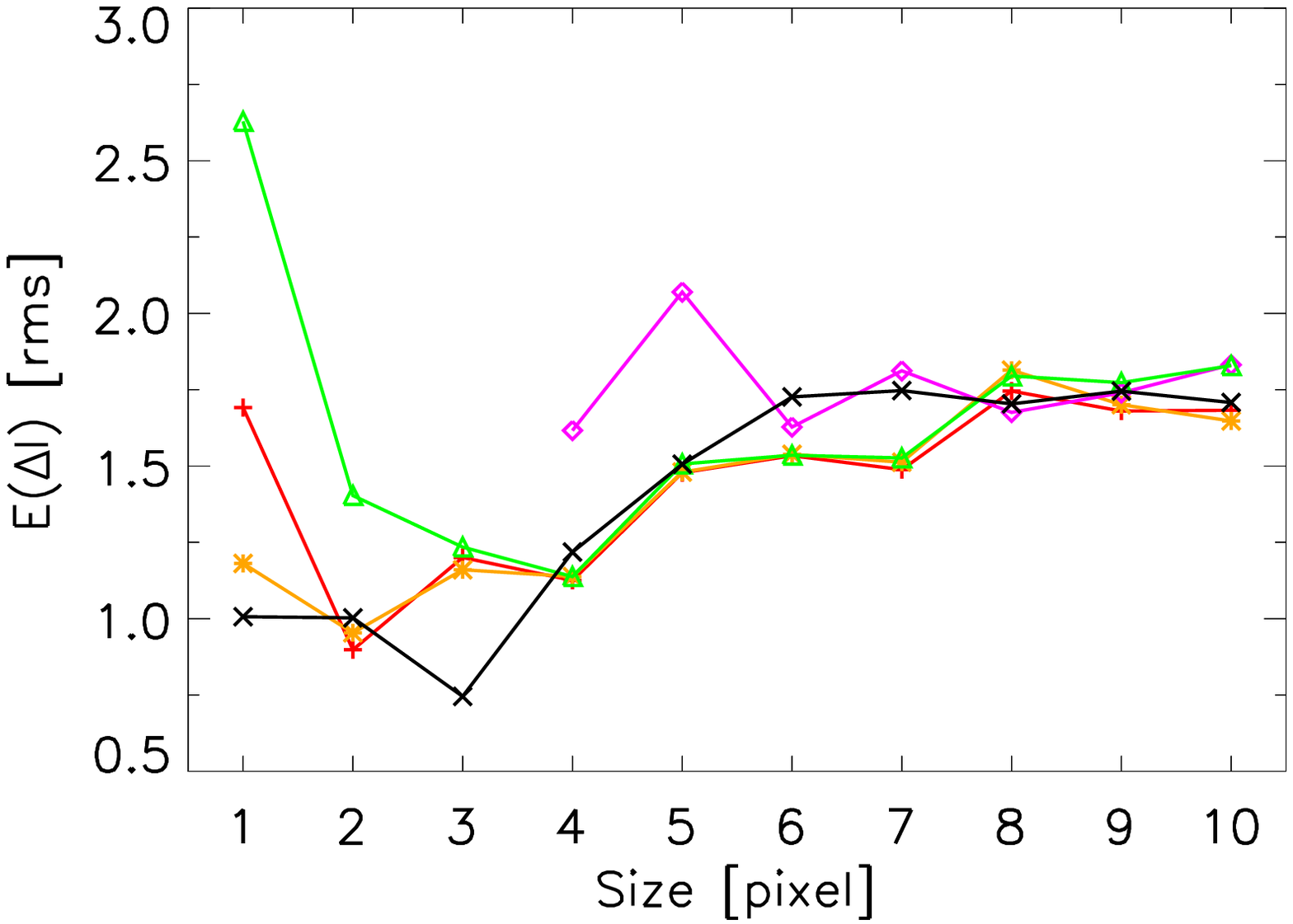}}
\subfloat[]{\label{fig:E_sum_size}
\includegraphics[width=0.4\textwidth]{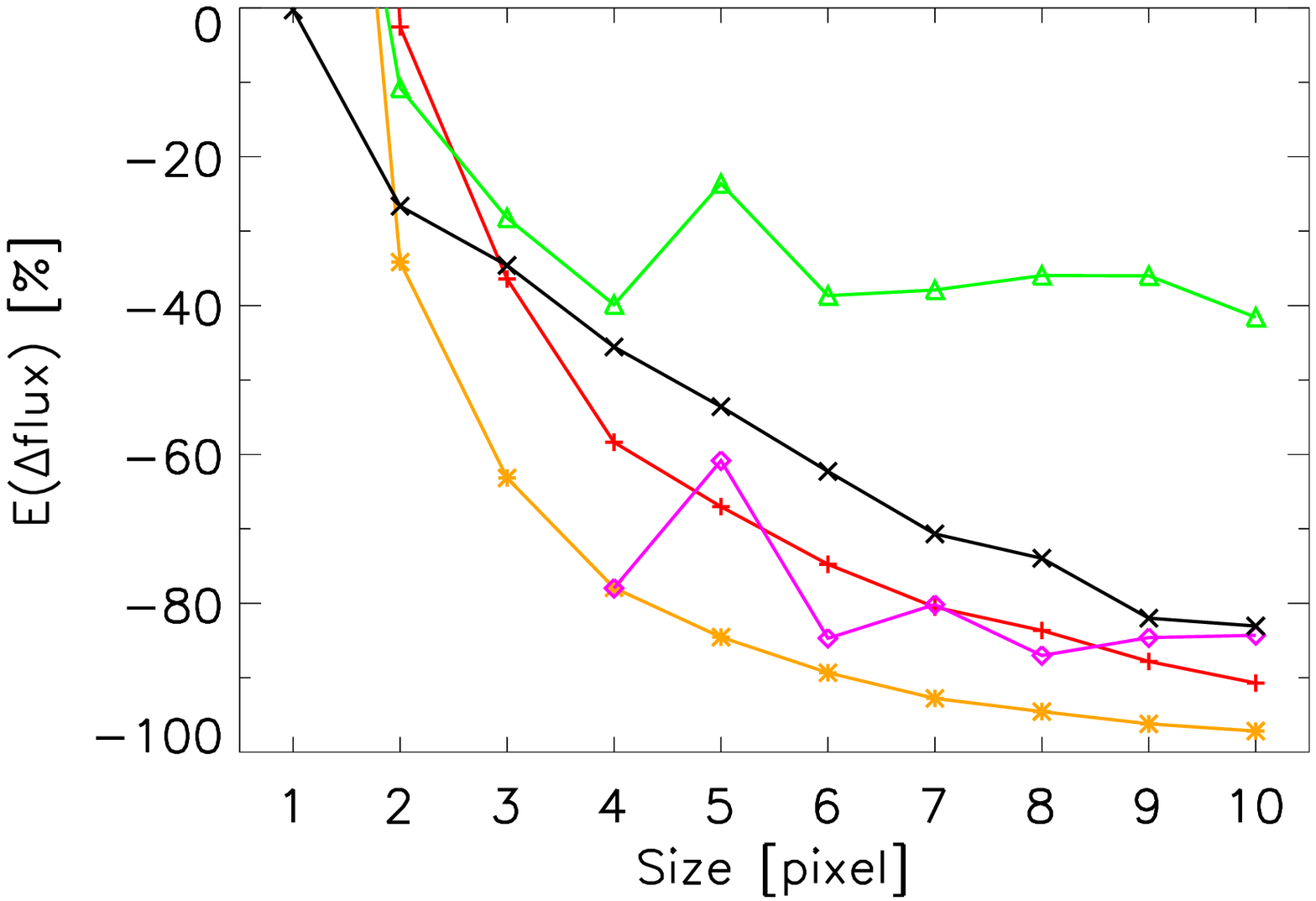}}
\caption{The mean error of position (a), size (solid line) and velocity dispersion (dash line) (b), peak brightness (c), and total flux (d) of algorithms in identifying clumps of different FWHM sizes.}
\label{fig:E_size}
\end{figure}

The average deviations in clump position, size, peak brightness, and the total flux as a function of input clump size are shown in Figure \ref{fig:E_size}. It can be seen that when the size of clumps are smaller than 5 pixels, the average error in clump position for Dendrograms is about 0.5 pixel, while the deviations for other algorithms are more than 0.5 pixels. The average deviation in clump position shows no trend with the size of clumps for all algorithms. The average errors in clump size and velocity dispersion for each algorithm gradually increase as the size of clumps increases. As shown in Figure \ref{fig:E_s_size}, the average error in clump velocity dispersion is generally less than the error in size for most of the algorithms. However, the relative errors of the size are similar to the errors of velocity dispersion (see Table \ref{tab3}). The largest errors in clump size, velocity dispersion, and peak brightness for all algorithms are about -9 pixels (ClumpFind2006), -3 pixels (Gaussclumps), and  2.5 times of the noise (Fellwalker), respectively. As the clump becomes larger, the error in peak brightness decreases to about 1.7 times the noise for all algorithms. Except for Fellwalker, all algorithms return clump sizes that are smaller than the input clump sizes. The clump peak brightness retrieved by all algorithms is higher than the input parameters. The total flux of a clump is the sum of brightness at all pixels within the boundary of the clump. For an optically thin clump, this parameter is proportional to the clump mass. As shown in Figure \ref{fig:E_sum_size}, Fellwalker exhibits the best performance compared to the other algorithms in the aspect of total flux, with about 60$\%$ of the total flux of the simulated clump being retrieved. For the other algorithms, the output total fluxes are lower than 40$\%$ of the simulated clumps. A reasonable explanation of this large error in clump total flux is the large error in the clump size, i.e., only a small fraction of the clump total flux is counted. Another possible reason is the omission of part of the flux by the algorithms. When the algorithms perform clump identification, the voxels with brightness below a designated value, which is adopted to be 3 times the noise level in our tests, are considered to be the noise and are ignored.


\subsection{Test 2: Performance of the Algorithms with Data of Different Signal-Noise Ratios}
\label{test2}

\begin{figure}[h]
  \centering
\includegraphics[width=0.3\textwidth]{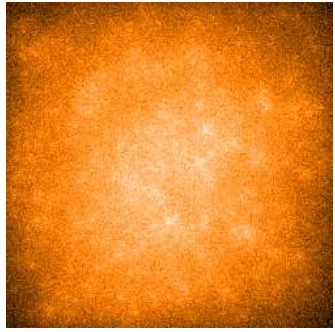}
  \caption{Simulated clumps with the signal-to-noise ratios distributing randomly from 1 to 21.}
  \label{fig:100cores_snr}
\end{figure}

In addition to the fact that the size of the input clumps has a significant impact on the clump identification results, the performance of an algorithm in identifying clumps with data of different peak brightness is also an important issue. In the 1000$\times$1000$\times$1000 array, we generated 1000 clumps with signal-to-noise ratios (SNR) distributing randomly from 1 to 21 (Figure \ref{fig:100cores_snr}). As in Section \ref{test1}, the six algorithms are used to search the simulated clumps to estimate the performance of different algorithms. We set the size of the clumps (FWHM) to be 5 pixels. In Section \ref{completeness1} it can be seen that when the size of clumps is large than 5 pixels, most of the algorithms perform well in the aspect of completeness. In current test the influence of size and crowdedness are reduced as much as possible. We focused on the performances of the algorithms when the SNRs of the data are changed.

\subsubsection{Completeness of the Algorithms}
\label{completeness2}

\begin{figure} [!htb]
\centering
\subfloat[]{\label{fig:right_snr} 
\includegraphics[width=0.4\textwidth]{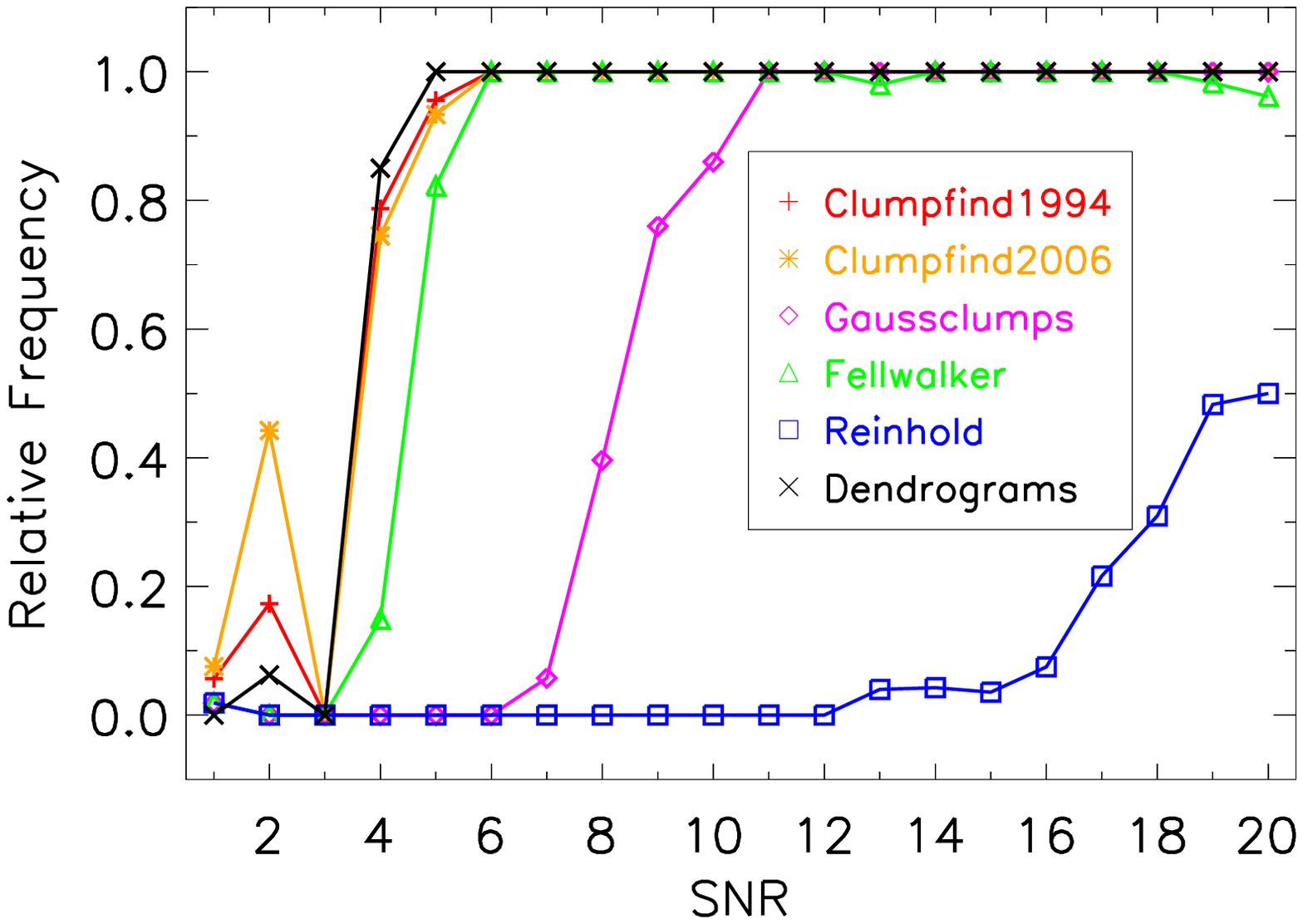}}
\subfloat[]{\label{fig:repeat_snr}
\includegraphics[width=0.4\textwidth]{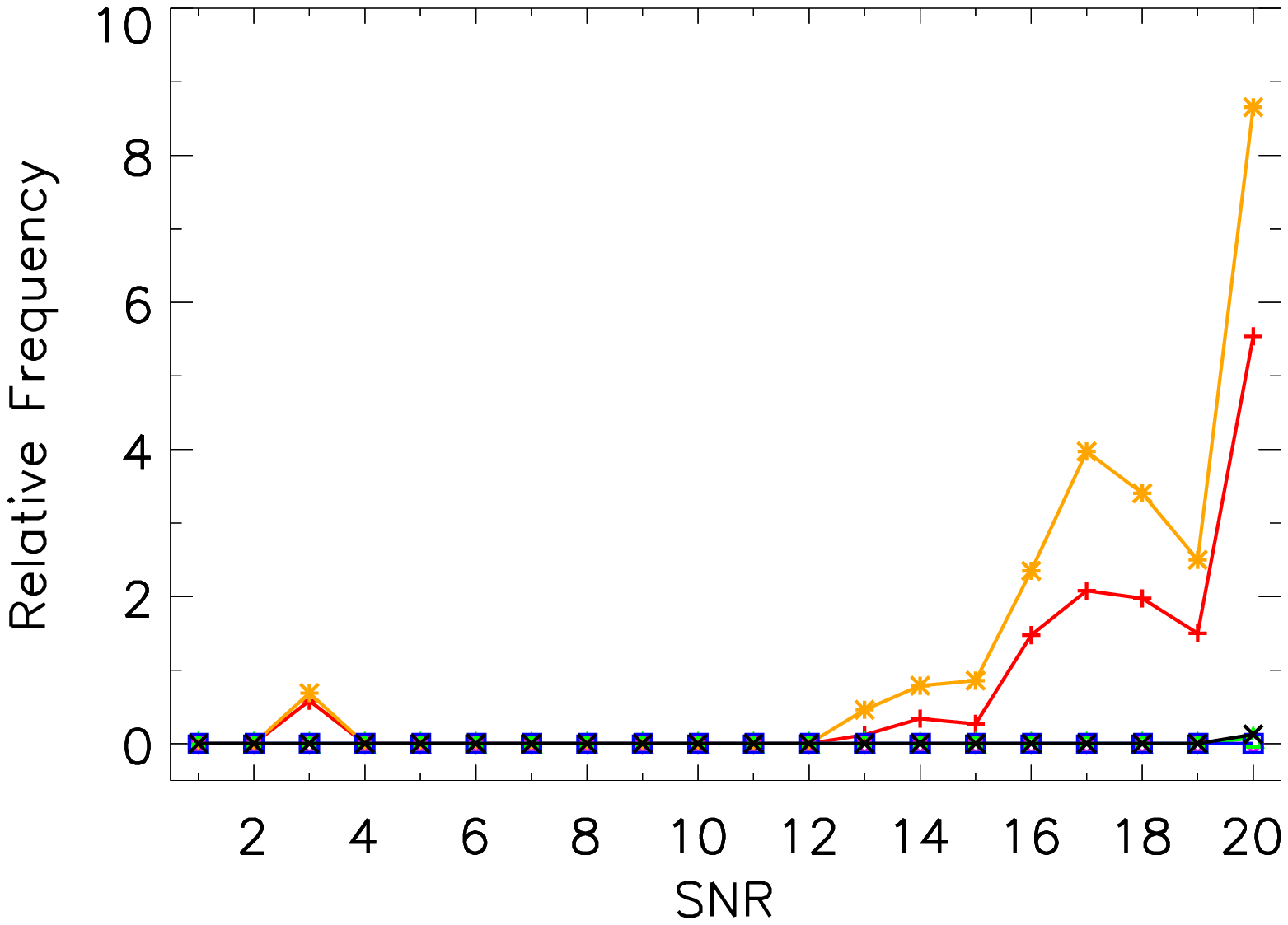}}\\
\subfloat[]{\label{fig:false_snr}
\includegraphics[width=0.4\textwidth]{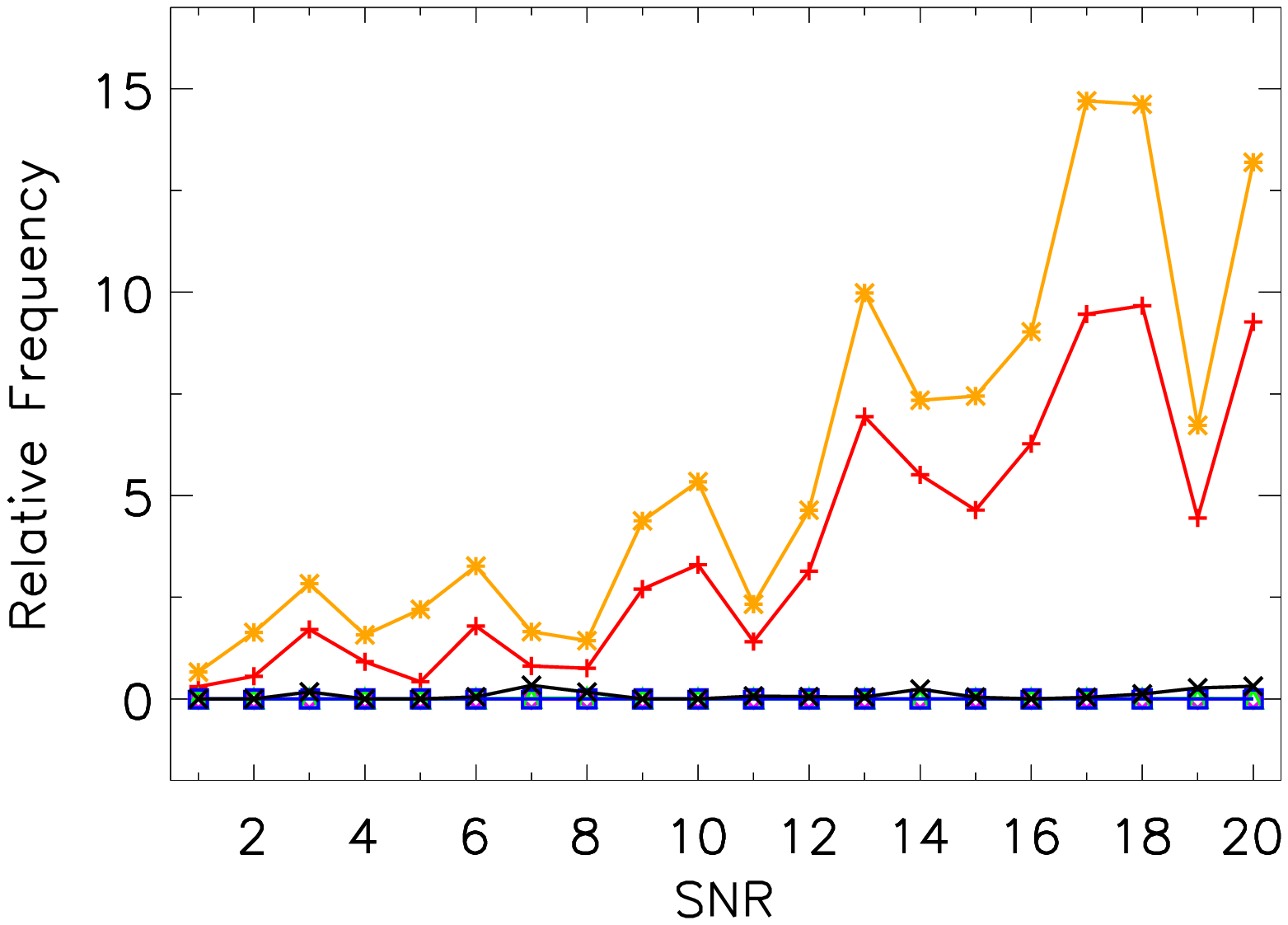}}
\subfloat[]{\label{fig:mark_snr}
\includegraphics[width=0.4\textwidth]{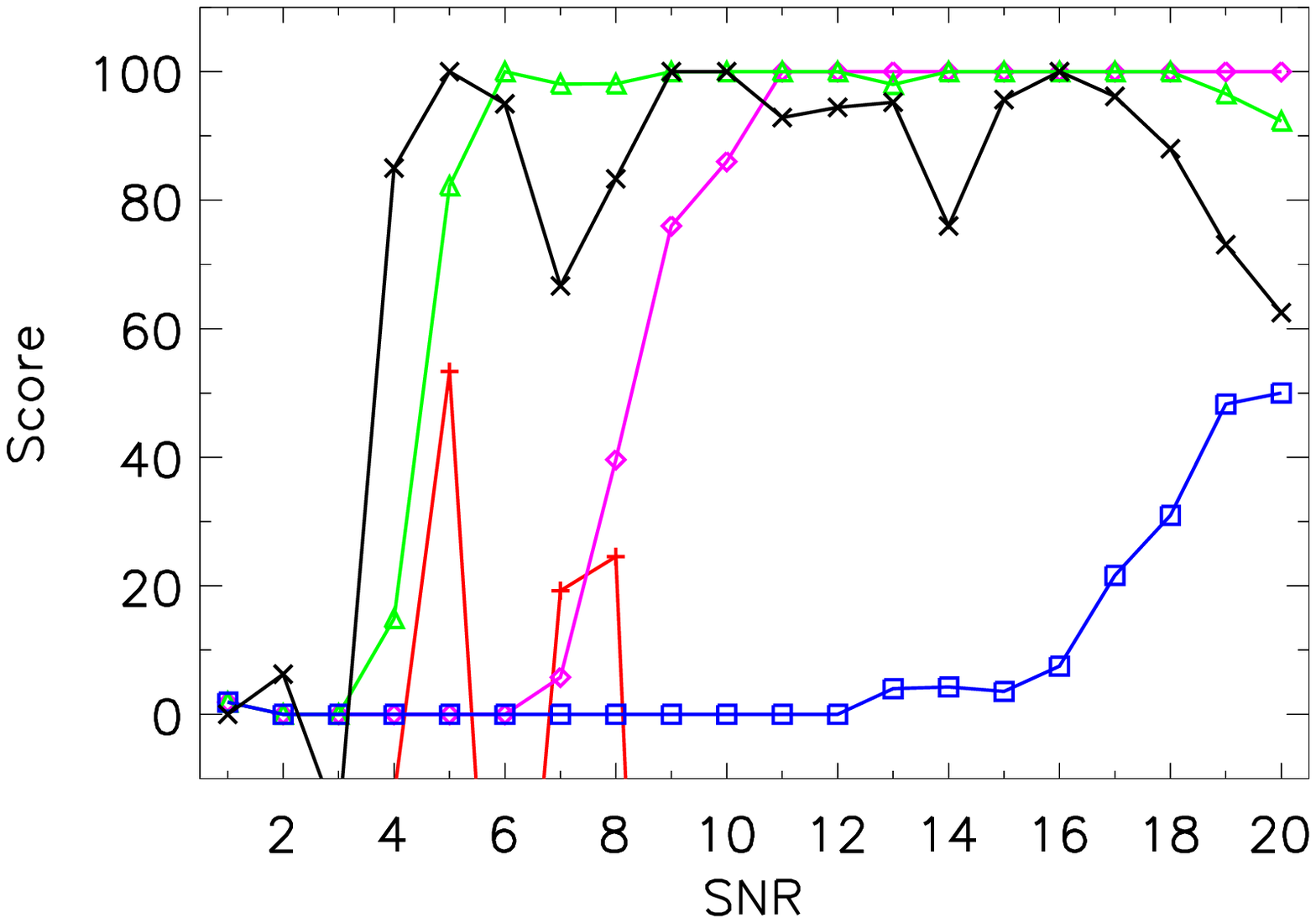}}
\caption{Frequeny distributions of the correct identifications (a), repeated identifications (b), erroneous identifications (c), and the score points (d) of the six algorithms in identifying clumps with data of different signal-noise ratios.}
\label{fig:true_repeat_snr}
\end{figure}

Figure \ref{fig:right_snr} shows the number of correct identifications for all algorithms. It can be seen that all algorithms cannot detect clumps with peak brightness around the noise level. When the SNR of clumps is less than 3, all algorithms exhibit poor performance. When the SNR reaches 5, the numbers of correct identifications for all the six algorithms except Gaussclumps and Reinhold are all higher than 75$\%$. Reinhold and Gaussclumps exhibit lower completeness than the other algorithms when SNR is lower than 5. However, as the SNR increases, the accuracy of Reinhold and Gaussclumps increases. The accuracy of Gaussclumps reaches more than 90$\%$ when the SNR reaches 11. The accuracy of Reinhold reaches 20$\%$ only when the SNR is as large as 17.  Therefore, Reinhold and Gaussclumps are only suitable for searching for clumps with high brightness. Figure \ref{fig:repeat_snr} shows the number of repeated identifications for different algorithms. The ClumpFind2006 and ClumpFind1994 algorithms exhibit the highest repetitive rates. Duplicate identifications of Fellwalker, Reinhold, and Gaussclumps are fewer than other algorithms.

The numbers of erroneous identifications of different algorithms are presented in Figure \ref{fig:false_snr}. Surprisingly, as the SNR of the simulated clumps increases, the numbers of false clumps returned by ClumpFind1994 and ClumpFind2006 increase. The other algorithms almost do not erroneously count the fluctuation of the noise as a clump. As in Section \ref{completeness1}, we establish a simple scoring mechanism to evaluate the overall performance of each algorithm. The scores of the algorithms are shown in Figure \ref{fig:mark_snr}. It can be seen that Fellwalker, Dendrograms, and Gaussclumps are the best algorithms. ClumpFind1994 and ClumpFind2006 score low due to their many false identifications.

\subsubsection{Accuracy of Retrieved Parameters}
\label{accuracy2}

\begin{figure} [!htb]
\centering
\subfloat[]{\label{fig:} 
\includegraphics[width=0.4\textwidth]{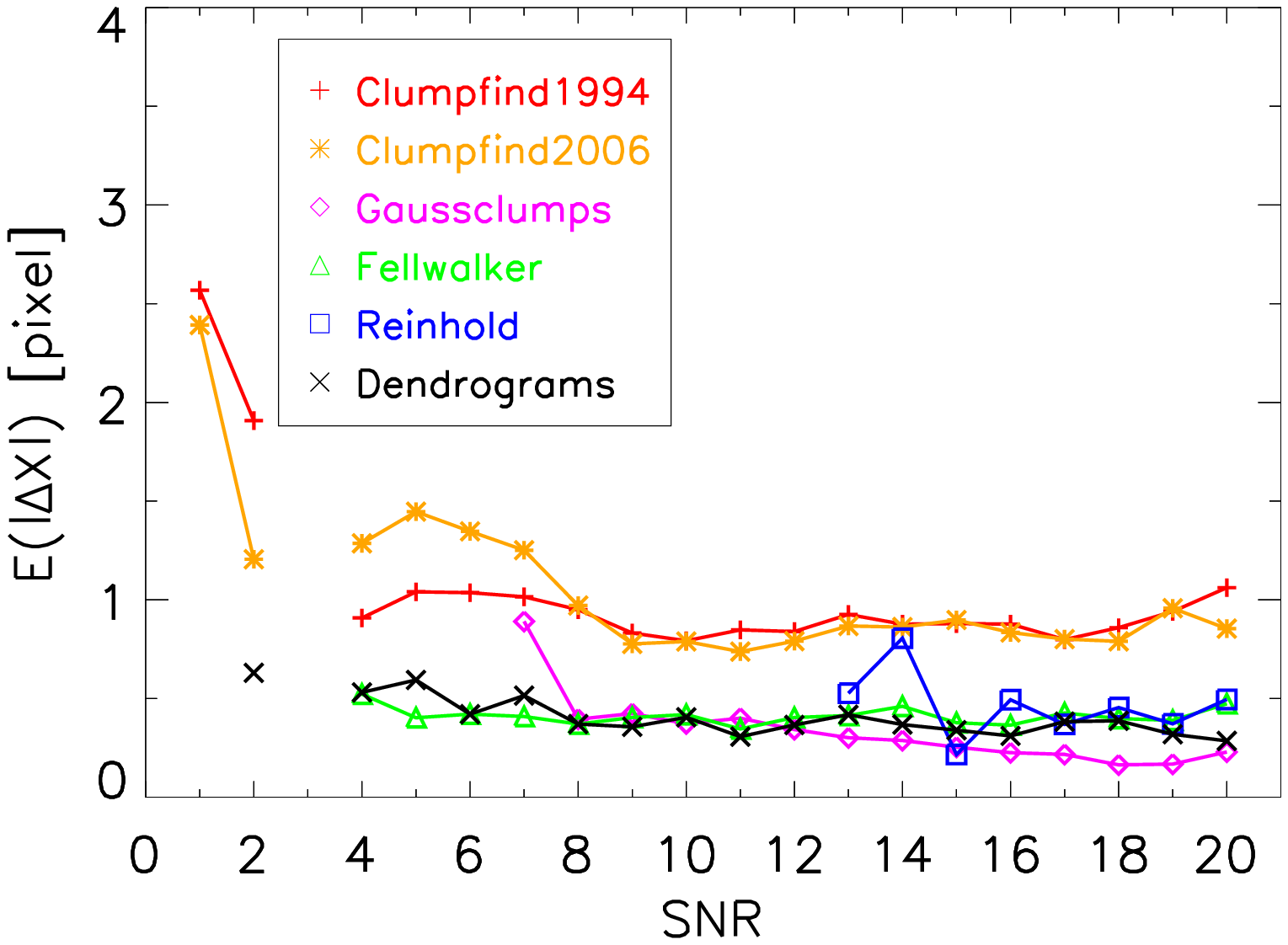}}
\subfloat[]{\label{fig:}
\includegraphics[width=0.4\textwidth]{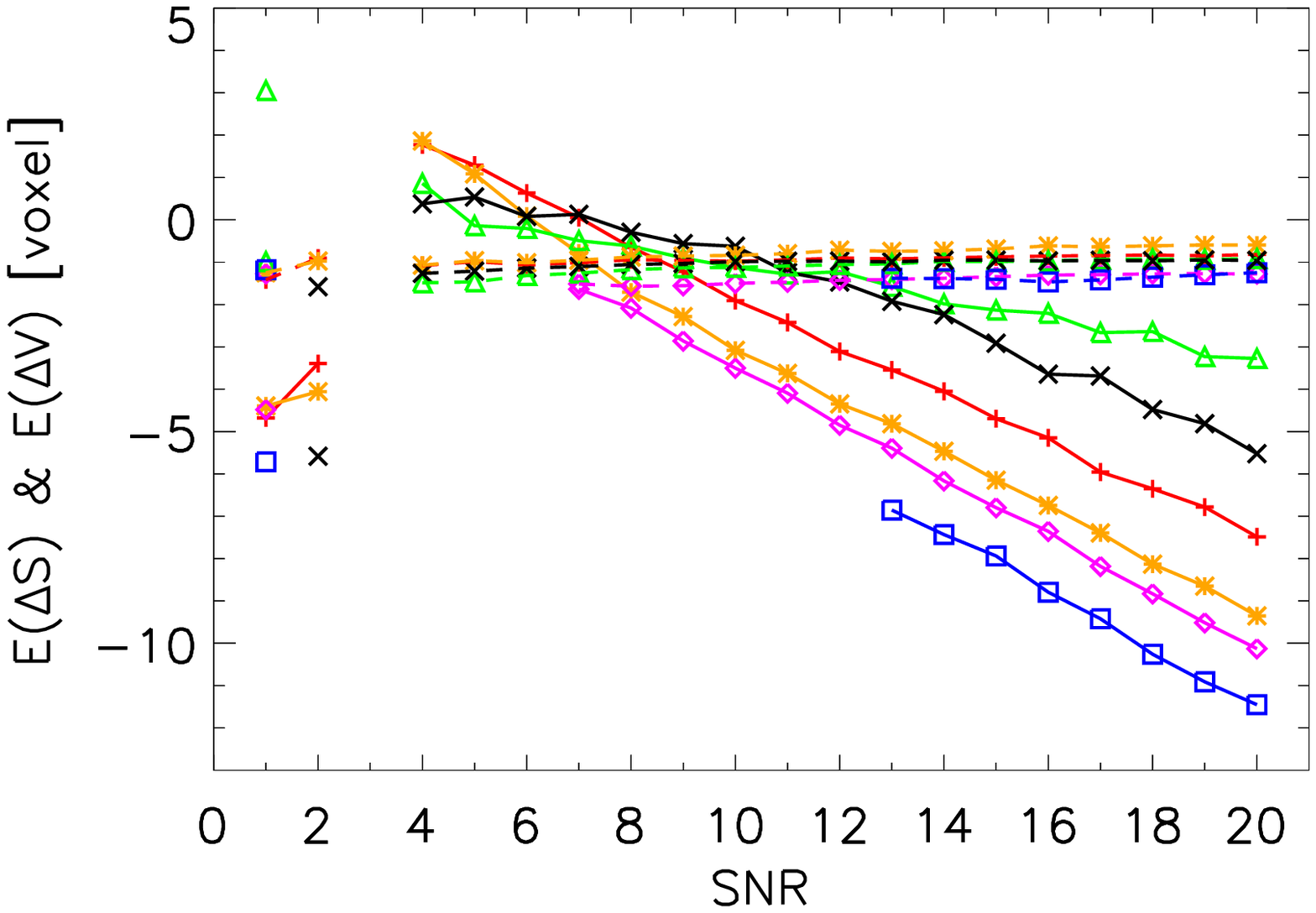}}\\
\subfloat[]{\label{fig:}
\includegraphics[width=0.4\textwidth]{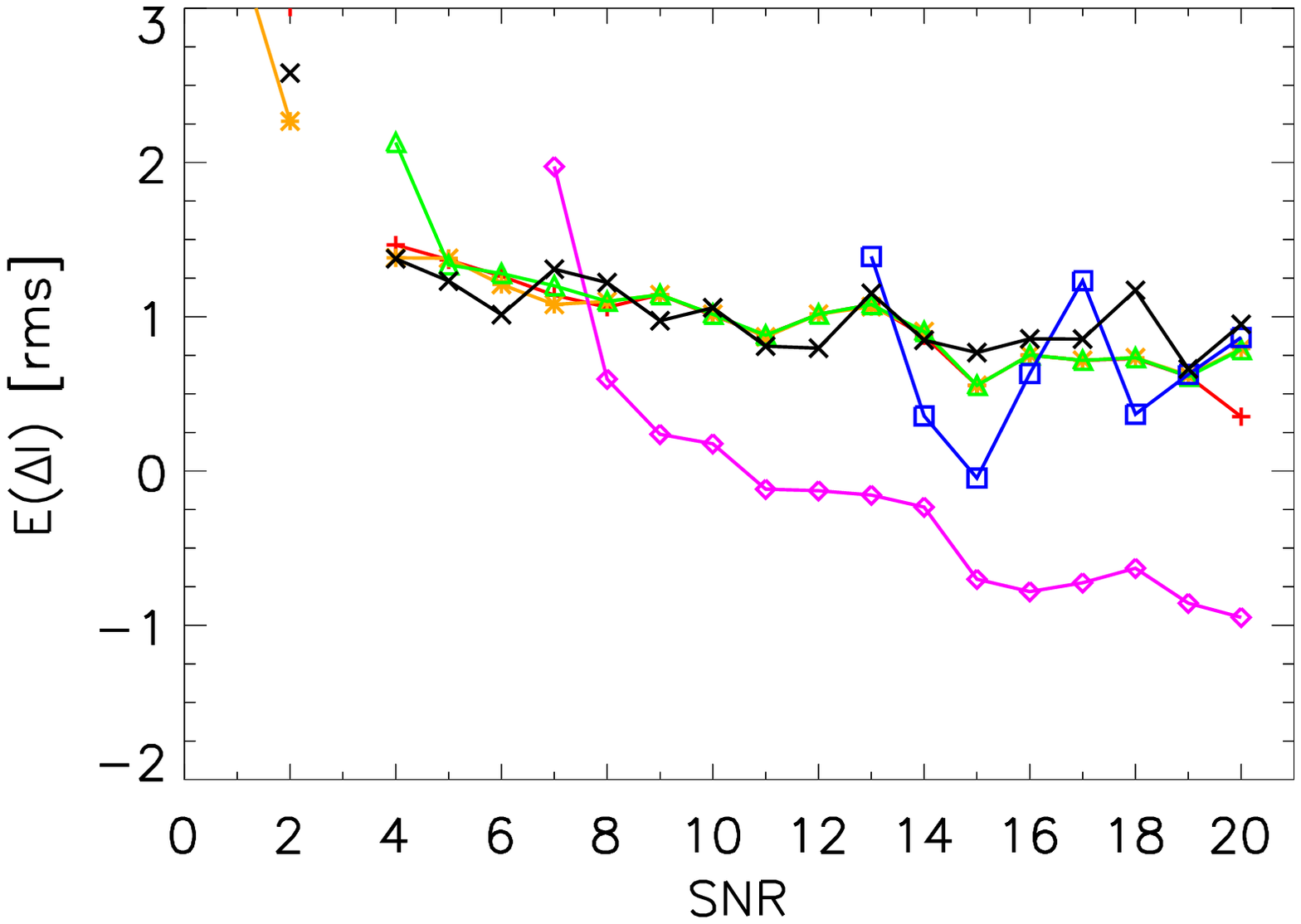}}
\subfloat[]{\label{fig:E_sum_snr}
\includegraphics[width=0.4\textwidth]{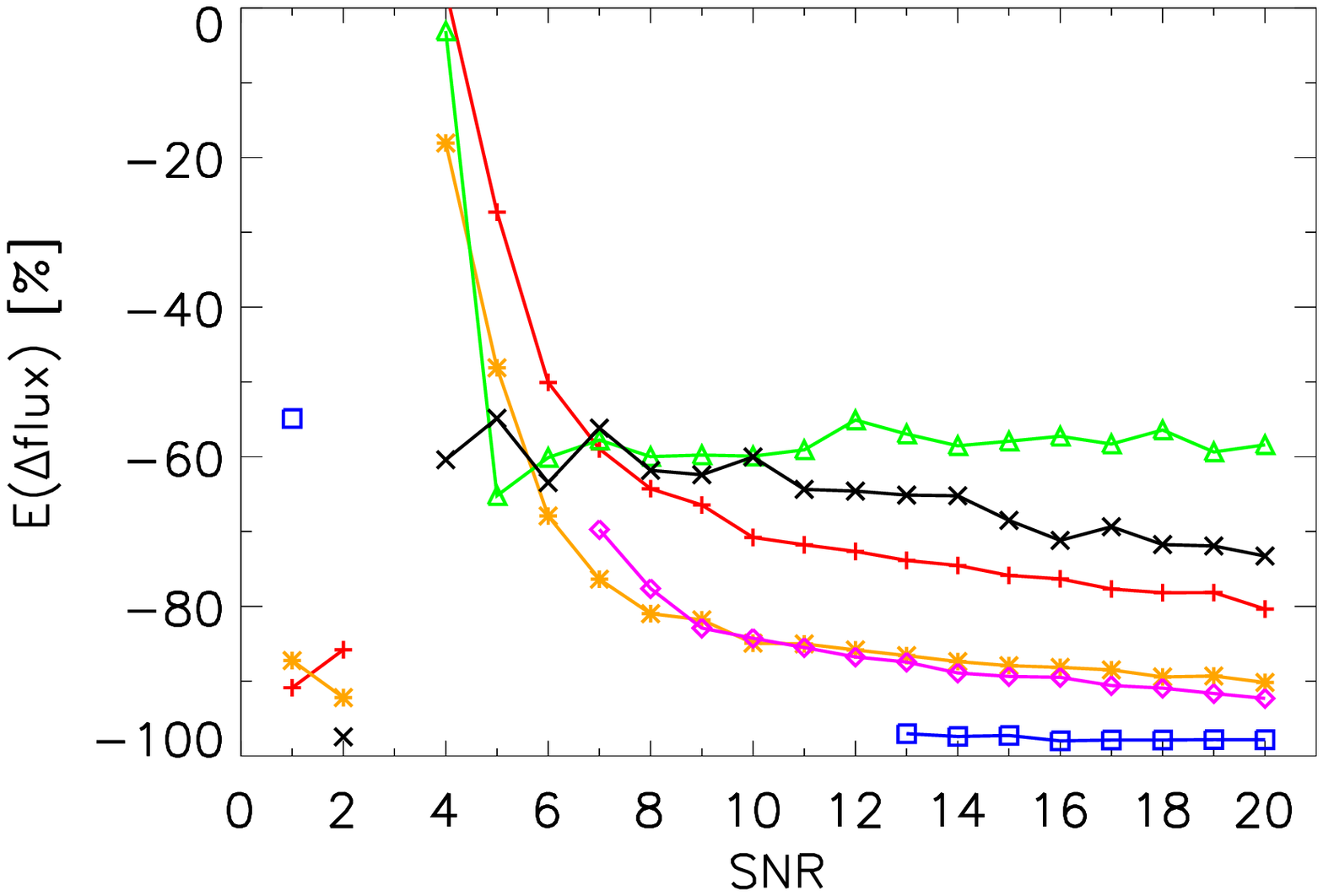}}
\caption{The average error of position (a), size (solid line) and velocity dispersion (dash line) (b), peak brightness (c), and total flux (d) of six algorithms in identifying clumps with data of different signal-to-noise ratios.}
\label{fig:E_snr}
\end{figure}

We calculated the average errors of the  position (E($|\Delta$X$|$)), size (E($\Delta$S)), velocity dispersion (E($\Delta$V)), peak brightness (E($\Delta$I)), and total flux (E(flux)) for each algorithm. These average errors as a function of different SNRs are shown in Figure \ref{fig:E_snr}. It can be seen that as the SNR of the clumps increases, the errors in the clump position and peak brightness gradually decrease. The average errors in clump size for each algorithm gradually increase as the SNR of clumps increases, while the average errors in clump velocity dispersion remain nearly constant. The overall performance of Fellwalker in reproducing the clump parameters is still good. As shown in Figure \ref{fig:E_sum_snr}, Fellwalker exhibit the best performance among the six algorithms in the aspect of total flux, with more than 40$\%$ of the total flux being extracted to output when the SNR reaches 20. Reinhold exhibits the biggest total flux deviation compared to the others. Fellwalker and Dendrograms perform better than other algorithms in retrieving the parameters. Most of the algorithms return a smaller size and velocity dispersion and a higher peak brightness than the simulated data, and the total fluxes of the output are always lower than the simulated clumps.

\subsection{Test 3: Performance of the Algorithms in Identifying Clumps with Different Crowdedness}
\label{test3}

\begin{figure}[h]
  \centering
\includegraphics[width=0.3\textwidth]{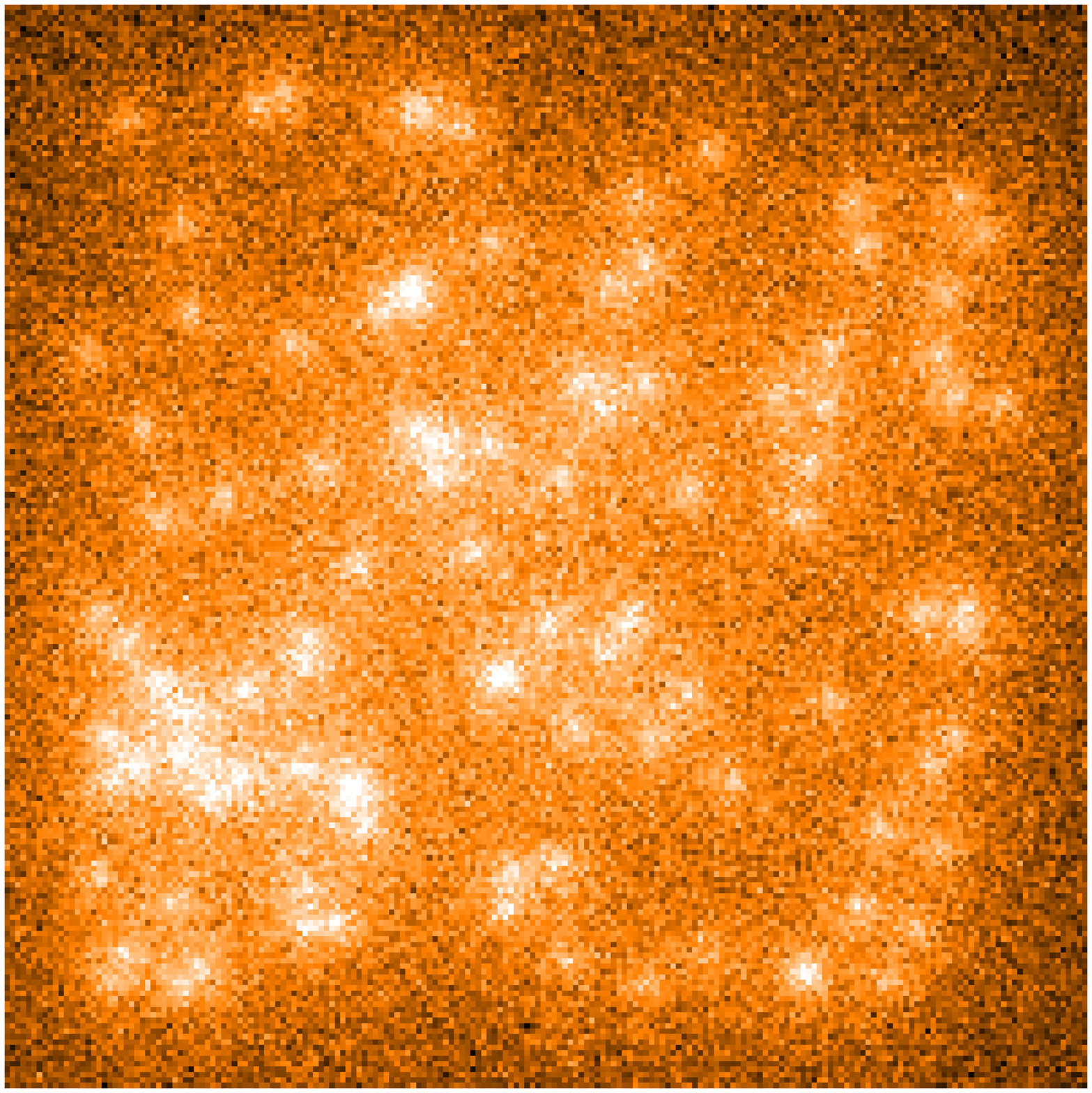}
\includegraphics[width=0.3\textwidth]{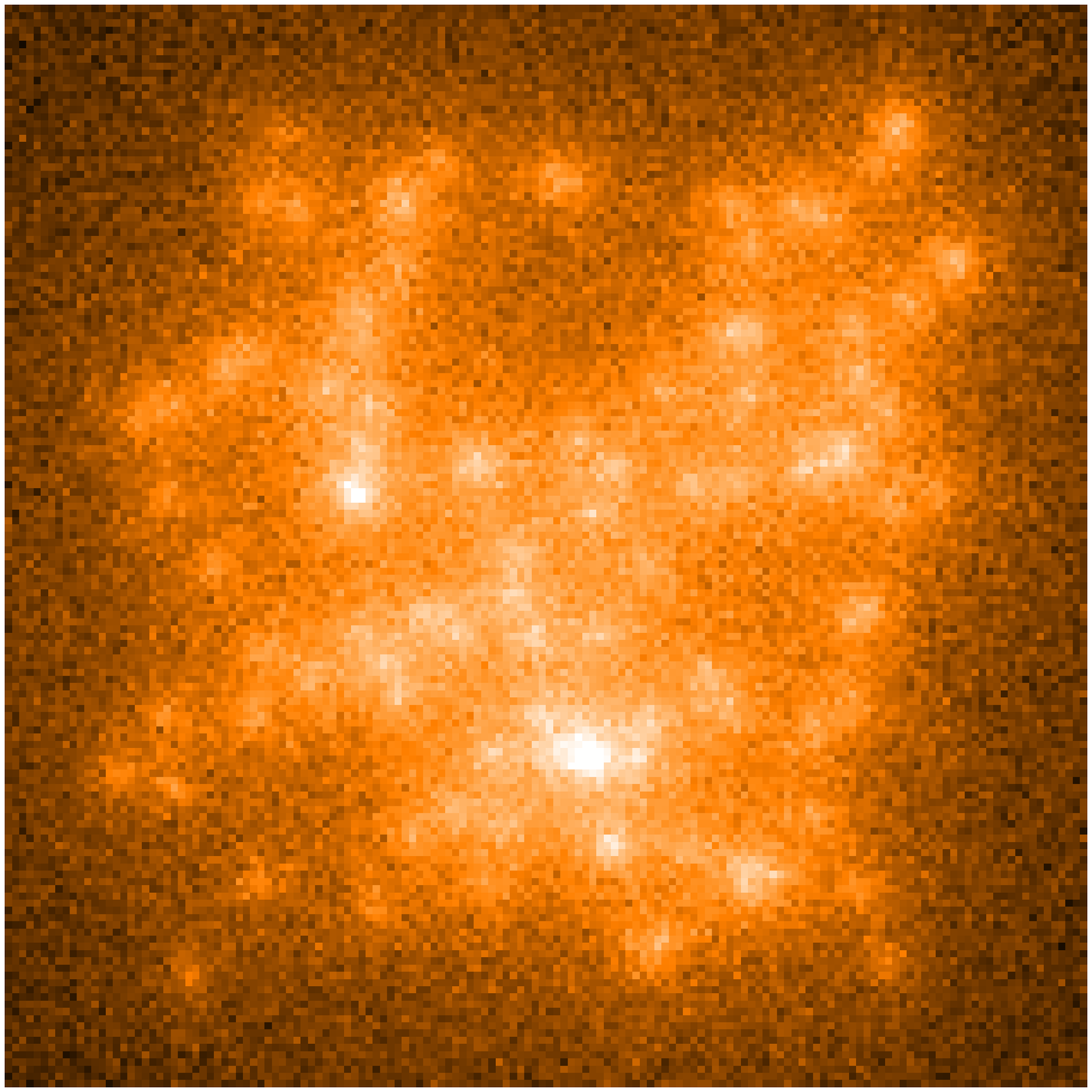}
\includegraphics[width=0.3\textwidth]{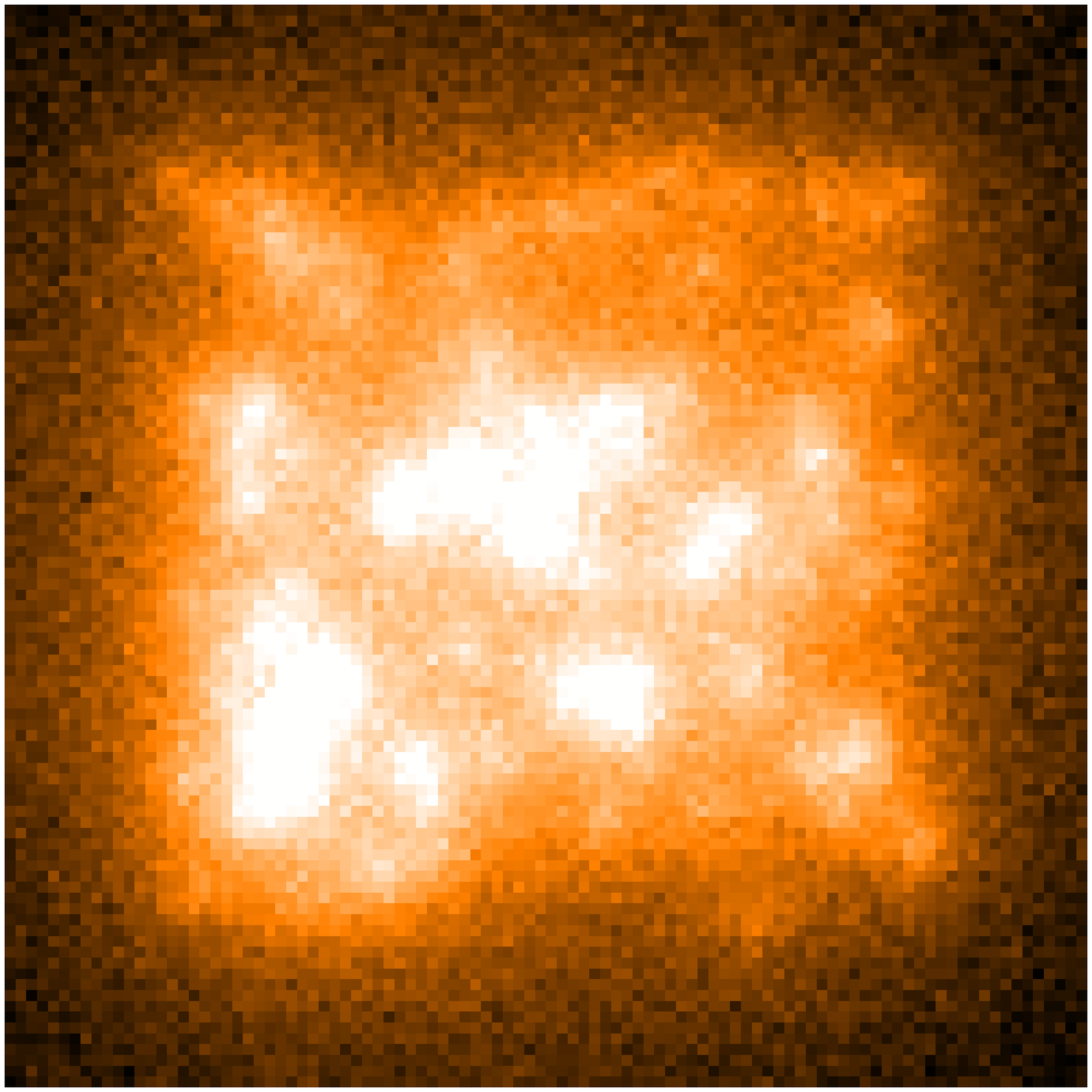}
  \caption{Distributions of 100 simulated clumps in 200$\times$200$\times$200 (left), 150$\times$150$\times$150 (middle), and 100$\times$100$\times$100 arrays (right), respectively.}
  \label{fig:100cores}
\end{figure}

It has long been realized that automated algorithms tend to interpolate clumps at various peak values and break the bright sources into multiple clumps \citep{2006ApJ...638..293E}. Therefore, the automated algorithms are susceptible to the crowdedness of the clumps. In order to investigate the performance of different algorithms in the identification of clumps with different crowdedness, we create 100 clumps in 200$\times$200$\times$200, 150$\times$150$\times$150, and 100$\times$100$\times$100 arrays, respectively (Figure \ref{fig:100cores}). In the 200$\times$200$\times$200 array, few clumps are overlapped. In the 150$\times$150$\times$150 array some clumps overlaps at their edges. In the most crowded case (100$\times$100$\times$100 ), many clumps overlap. We identify the clumps with the six algorithms so that their performance can be unbiasedly estimated. 

We set the peak brightness of the clump to be 10 times the noise value, and the size to be 5 pixels. In this case, the influences of brightness and size are reduced as much as possible. We focus on the performance of the algorithms with different crowdedness. The completeness and accuracy of retrieved parameters from all the six algorithms are presented in Section \ref{completeness3} and \ref{accuracy3}.

\subsubsection{Completeness of the Algorithms}
\label{completeness3}

\begin{figure} [!htb]
\centering
\subfloat[]{\label{fig:right} 
\includegraphics[width=0.4\textwidth]{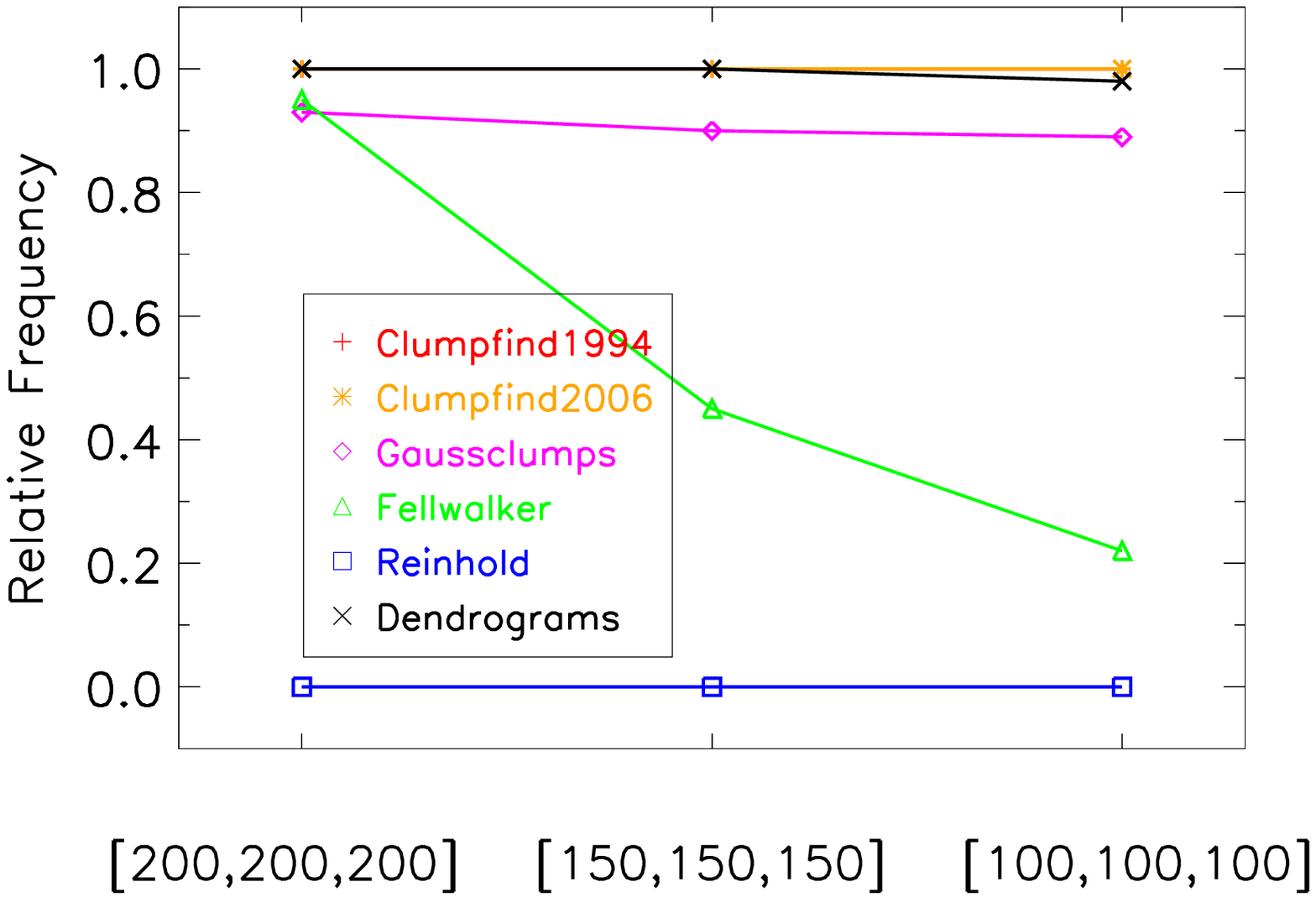}}
\subfloat[]{\label{fig:repeat}
\includegraphics[width=0.4\textwidth]{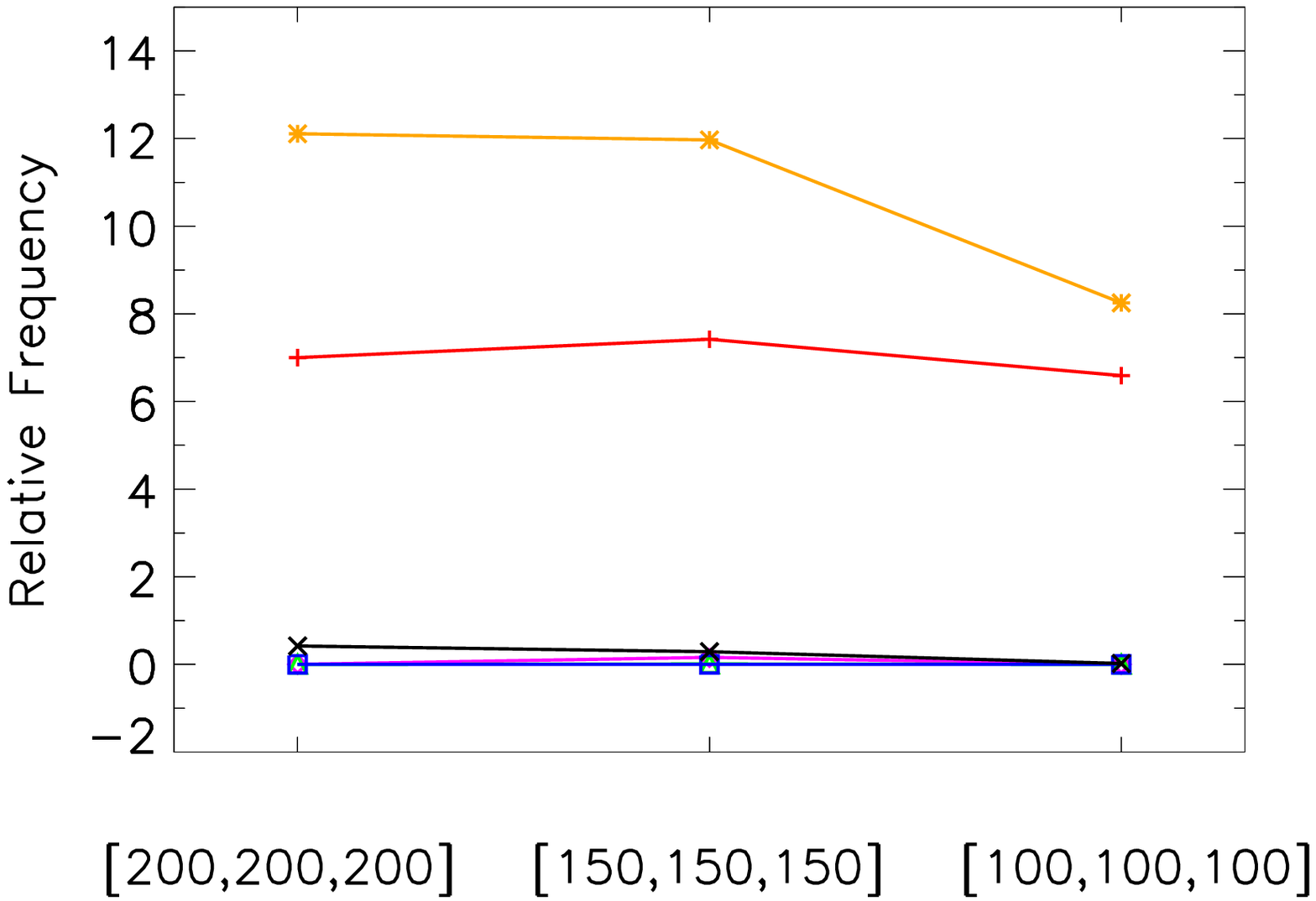}}\\
\subfloat[]{\label{fig:false}
\includegraphics[width=0.4\textwidth]{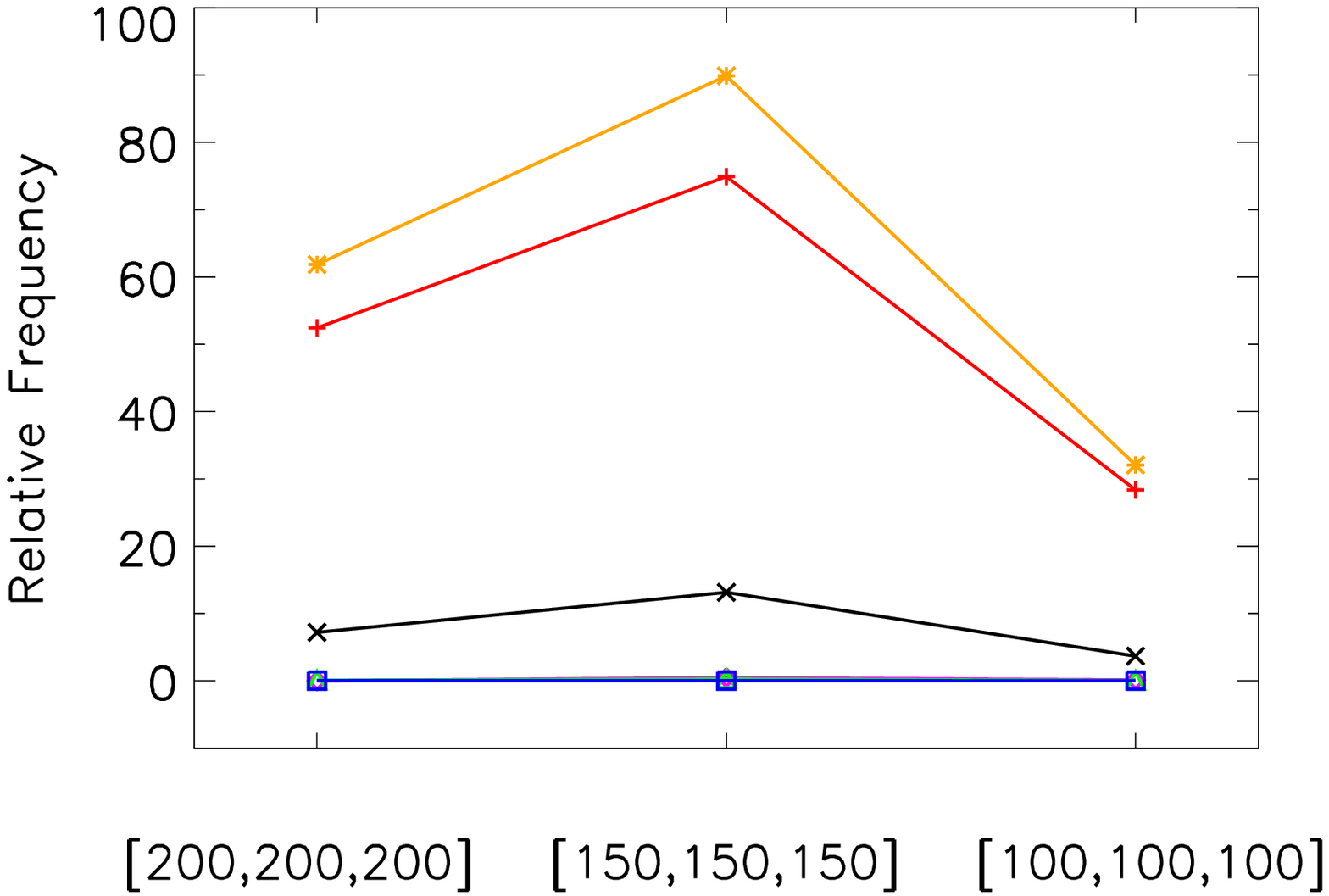}}
\subfloat[]{\label{fig:mark}
\includegraphics[width=0.4\textwidth]{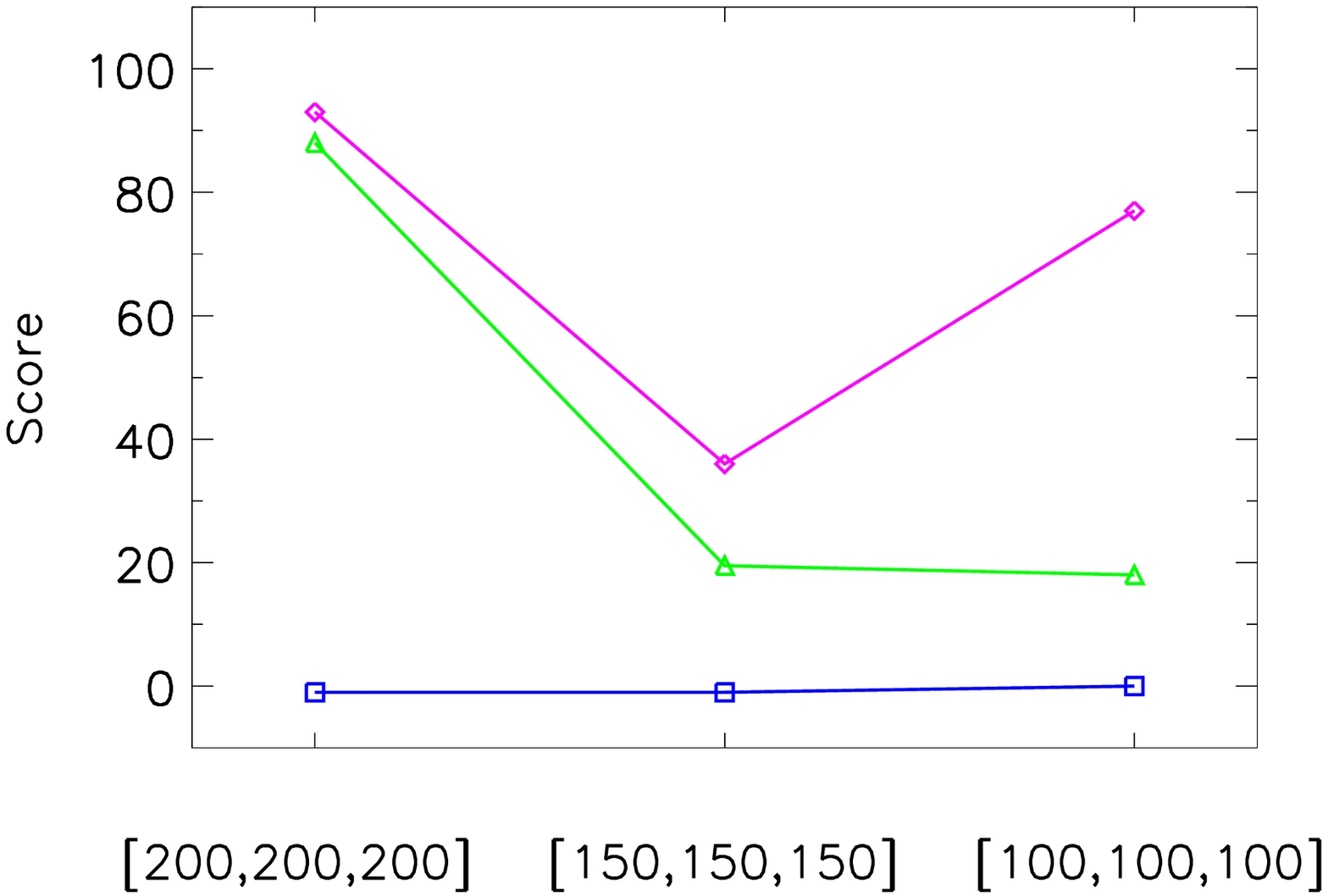}}
\caption{Frequency distributions of the correct (a), repeated (b), erroneous identifications (c), and the score points (d) of the six algorithms in identifying clumps with different crowdedness.}
\label{fig:true_repeat}
\end{figure}

Figure \ref{fig:right} shows the numbers of correct identifications of different algorithms. It can be seen that as the clumps become more crowded, the completeness of Fellwalker gradually decreases. ClumpFind1994, ClumpFind2006, Dendrograms, and GaussClumps perform better than Reinhold and Fellwalker in the aspect of completeness. In the case of [150, 150, 150], some clumps overlap at their edges and the brightness and boundary of the clumps are influenced by the ambient clumps. Accordingly, ClumpFind1994, ClumpFind2006, Dendrograms, and GaussClumps get more output numbers, repeated, and erroneous identifications than in the [200, 200, 200] arrays. Due to that many clumps overlap or merge into a new clump in the most crowded ([100, 100, 100]) case, the accuracy of all the algorithms fall down to $20\%-100\%$. 

Figure \ref{fig:repeat} shows the numbers of repeated identifications of different algorithms in the detection of clumps. Although ClumpFind1994 and ClumpFind2006 present good performance in accuracy, at the same time they repeatedly identified $600\%-1200\%$ clumps. This may be due to the fact that the ClumpFind algorithm tends to break the bright sources into multiple clumps \citep{2006ApJ...638..293E}. Fellwalker does not produce duplicate clumps, which is an important advantage compared to other algorithms.

Figure \ref{fig:false} shows the numbers of erroneous identifications of different algorithms. It can be seen that $2000\%-9000\%$ false clumps are outputted from ClumpFind1994 and ClumpFind2006. Fellwalker almost never identifies the fluctuation of the noise to be a clump.

\subsubsection{Accuracy of Retrieved Parameters}
\label{accuracy3}

\begin{figure} [!htb]
\centering
\subfloat[]{\label{fig:right_threshold} 
\includegraphics[width=0.4\textwidth]{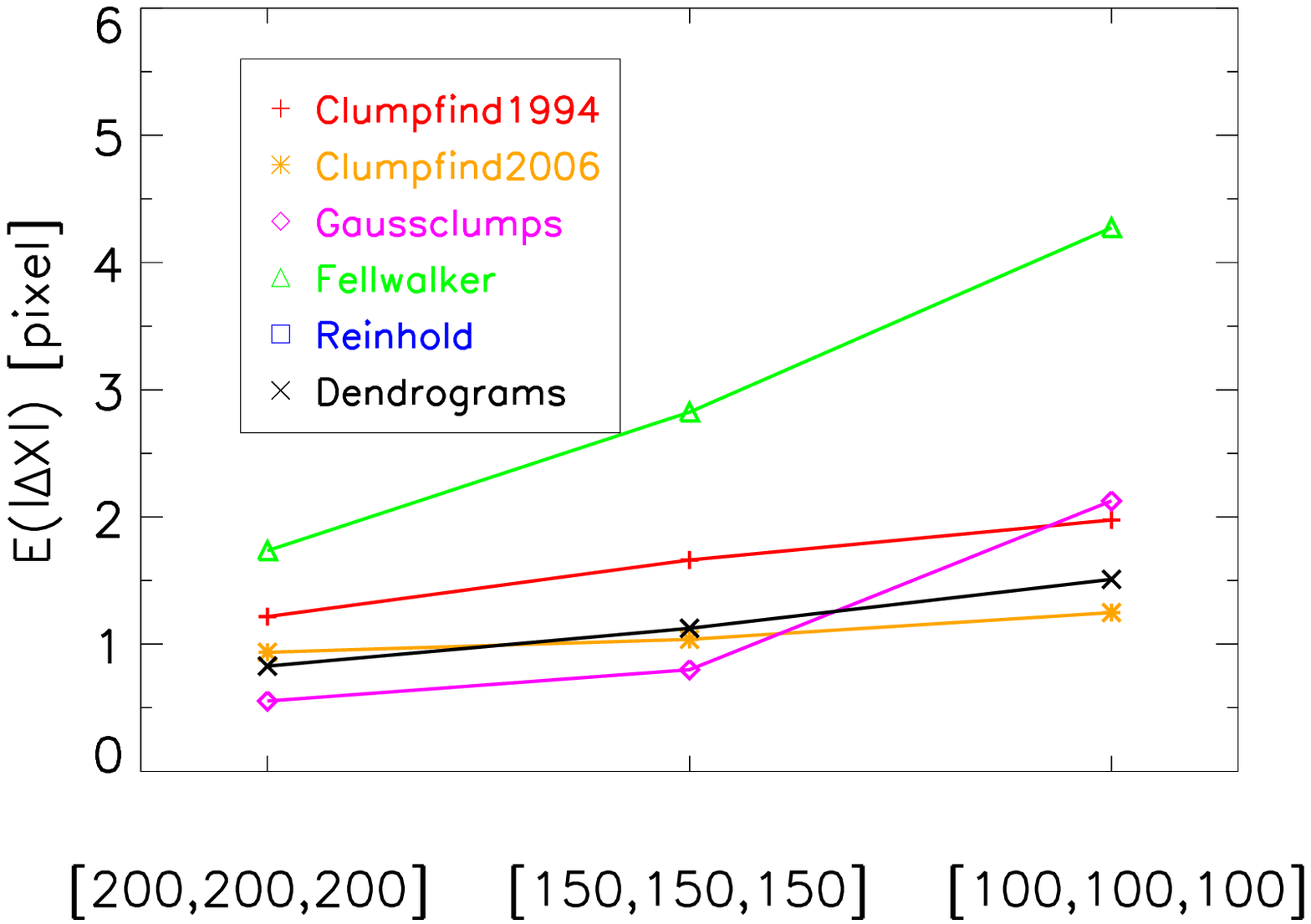}}
\subfloat[]{\label{fig:repeat_threshold}
\includegraphics[width=0.4\textwidth]{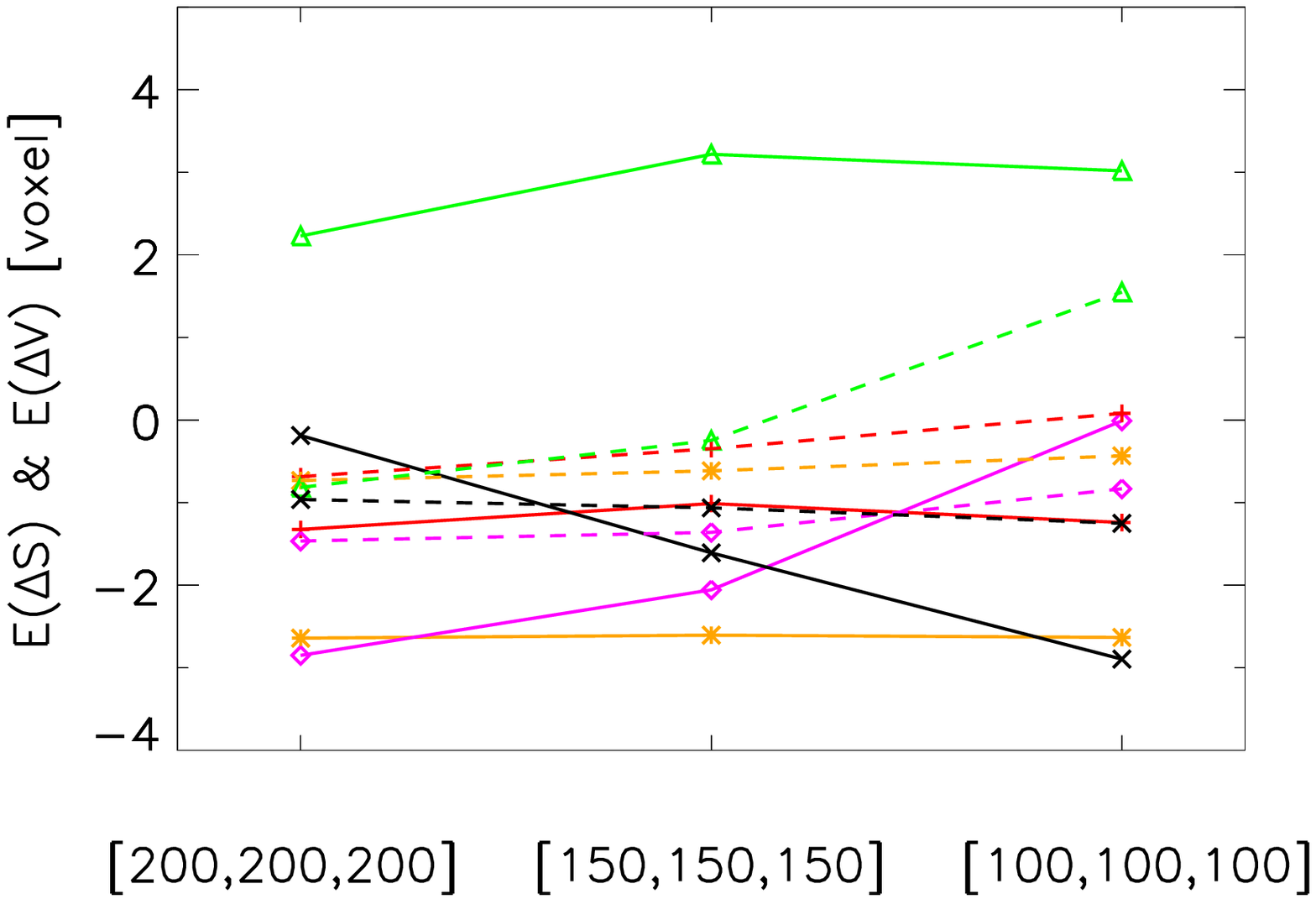}}\\
\subfloat[]{\label{fig:false_threshold}
\includegraphics[width=0.4\textwidth]{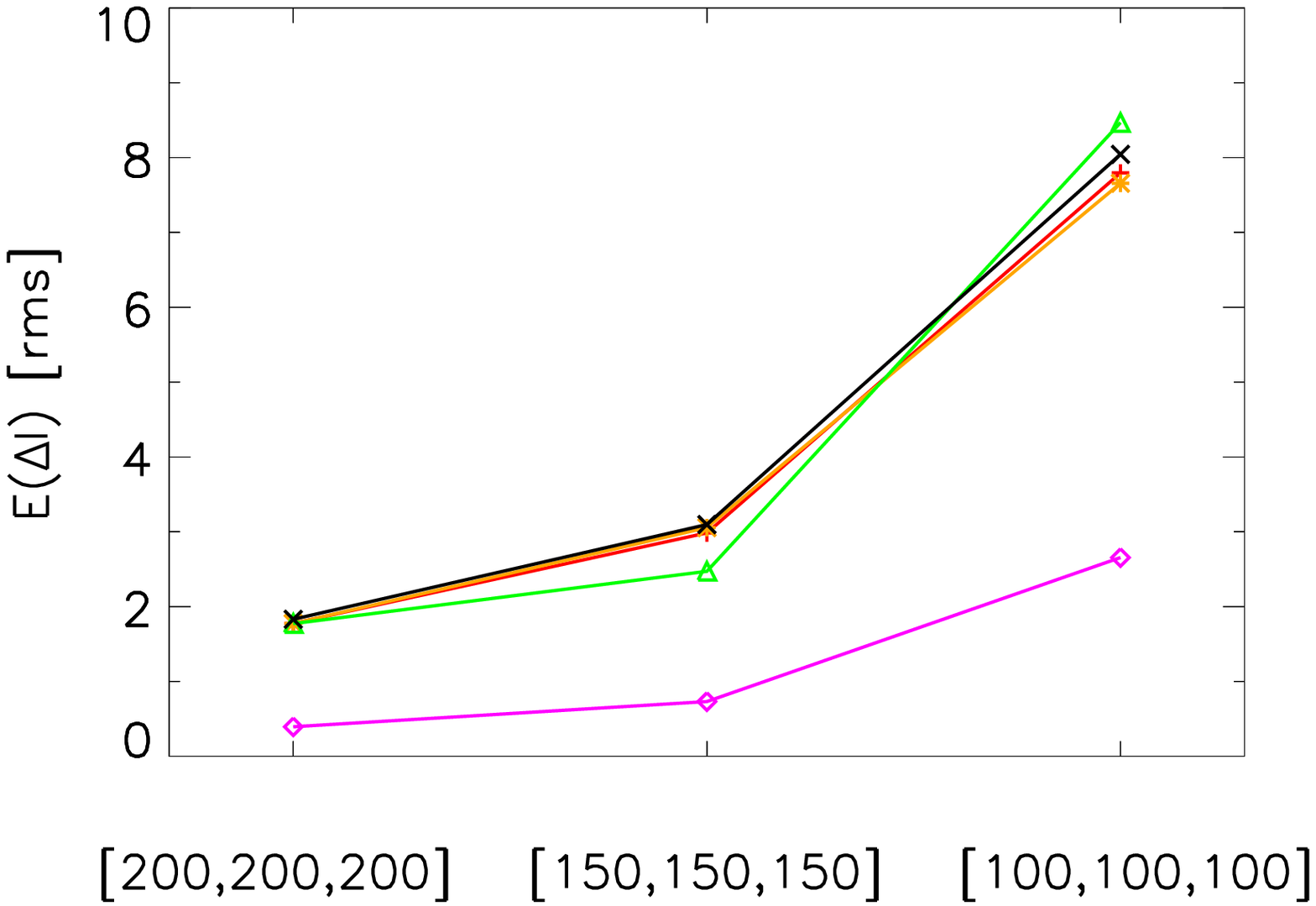}}
\subfloat[]{\label{fig:mark_threshold}
\includegraphics[width=0.4\textwidth]{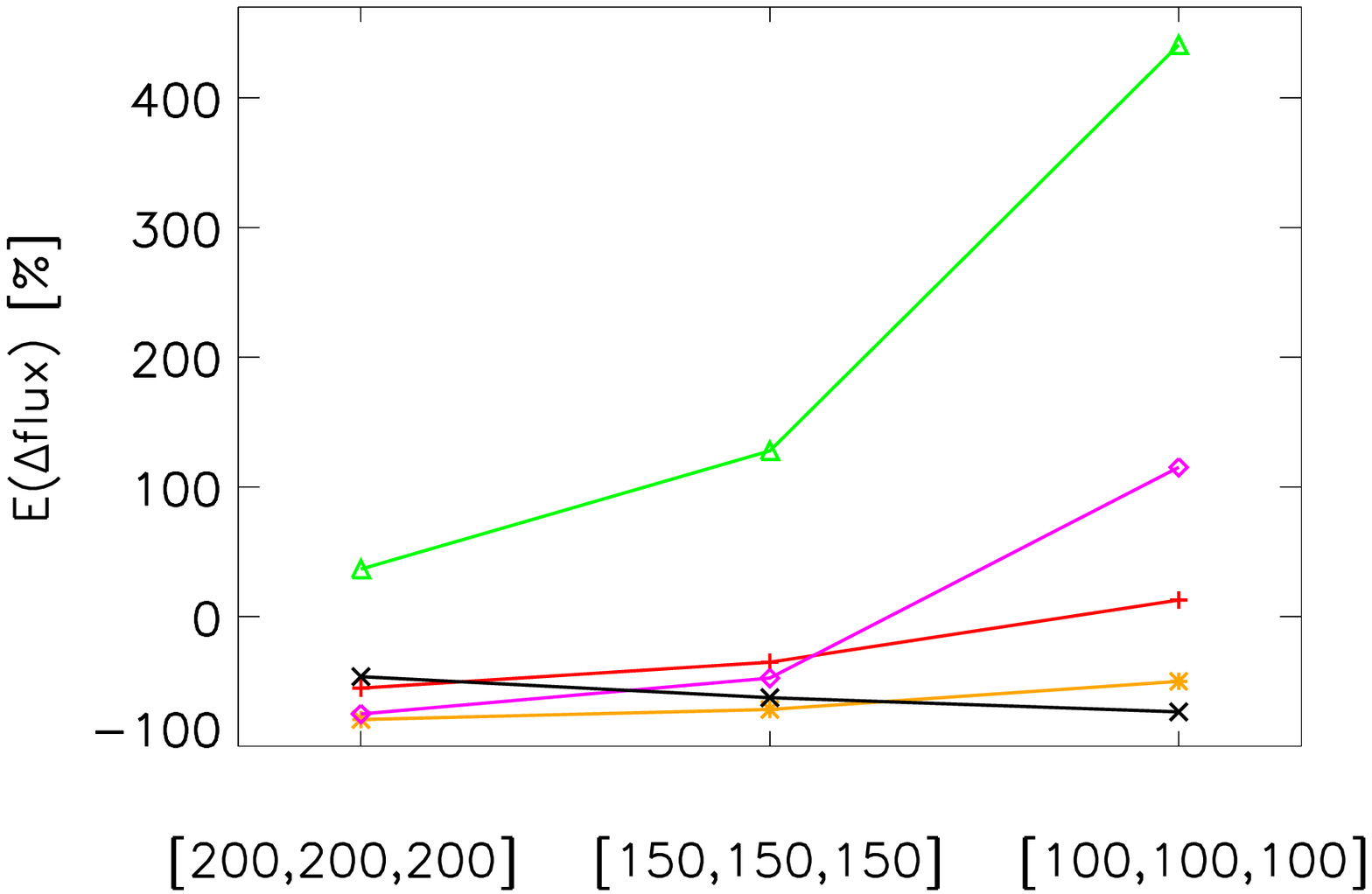}}
\caption{The average error of position (a), size (solid line) and velocity dispersion (dash line) (b), peak brightness (c), and total flux (d) of six algorithms in identifying clumps with different crowdedness.}
\label{fig:E}
\end{figure}

The average deviations of the corresponding retrieved parameters from the six algorithms in different crowdedness are shown in Figure \ref{fig:E}. It can be seen that in the most sparse case ([200, 200, 200]), the best algorithm for extracting the position parameter is Gaussclumps, with Dendrograms and ClumpFind2006 being the next. As the clumps get more crowded, the average errors in position and peak brightness gradually increase. Gaussclumps performs well in size and intensity extraction. For the peak brightness parameter, all the algorithms return values higher than the simulated data. The largest deviation in peak brightness among the six algorithms is about  8 times the one sigma noise which occurs in the most crowded case. In the most crowded case ([100, 100, 100]), the output total flux of Fellwalker, Gaussclumps, and ClumpFind1994 exceeds the input data, which may be caused by the merge of multiple clumps into a new clump.

\section{Discussion}
\label{discussion}

\subsection{The Bias of the the Algorithms in Estimation of the Virial Parameter}
\label{discussion1}

\begin{figure} [!htb]
\centering
\subfloat[]{\label{fig:vir_size} 
\includegraphics[width=0.4\textwidth]{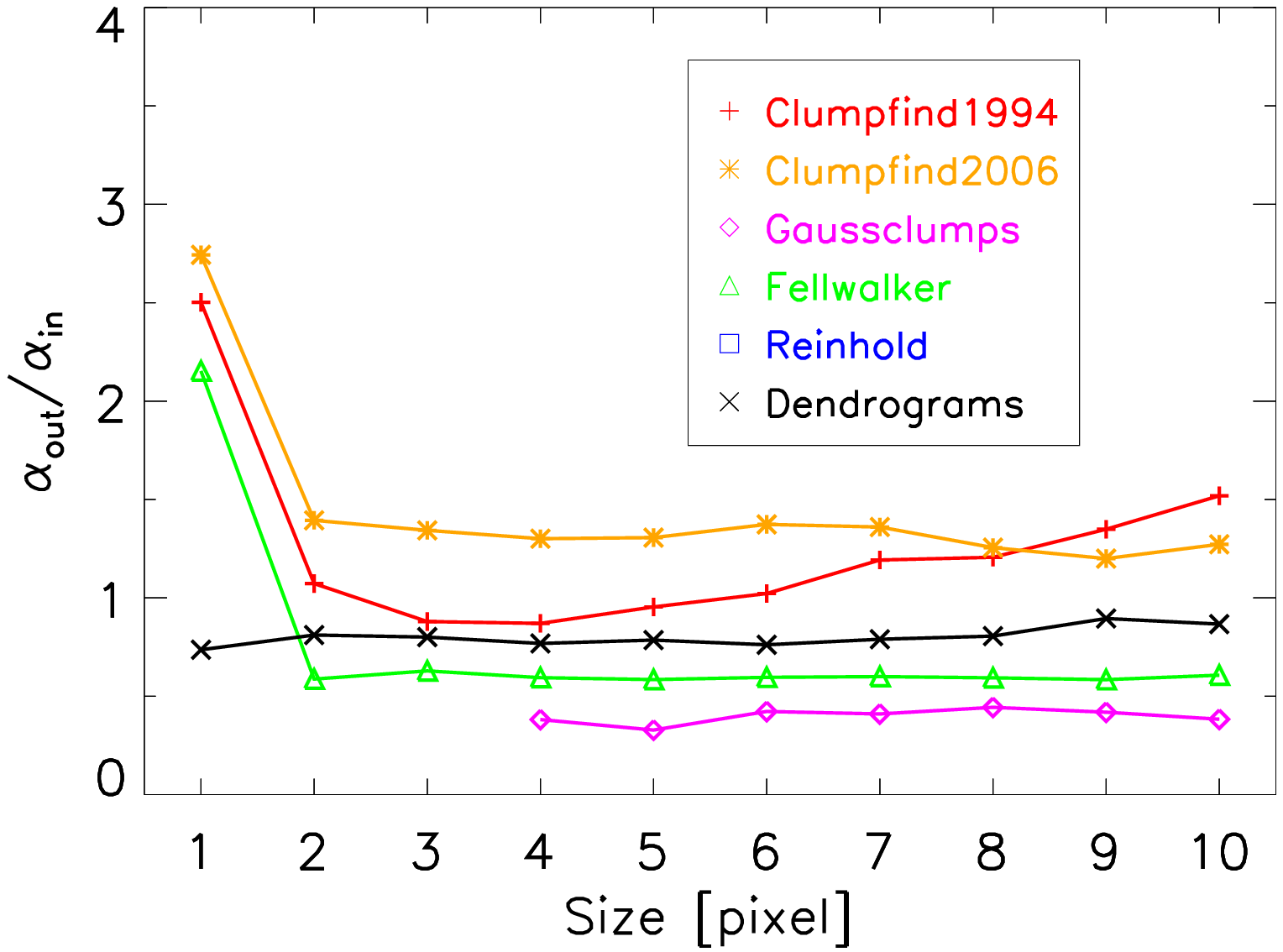}}
\subfloat[]{\label{fig:vir_snr}
\includegraphics[width=0.4\textwidth]{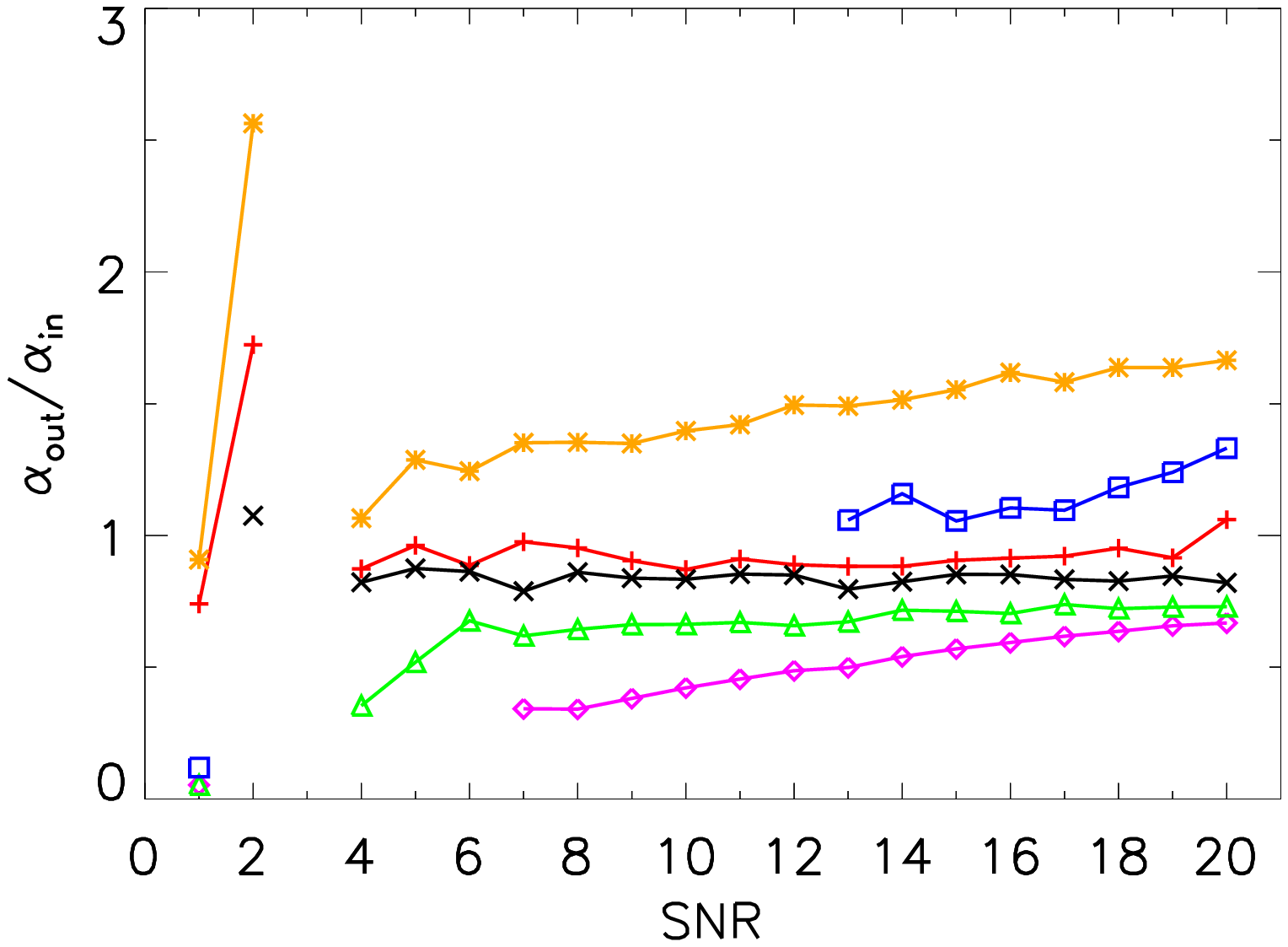}}\\
\subfloat[]{\label{fig:vir}
\includegraphics[width=0.4\textwidth]{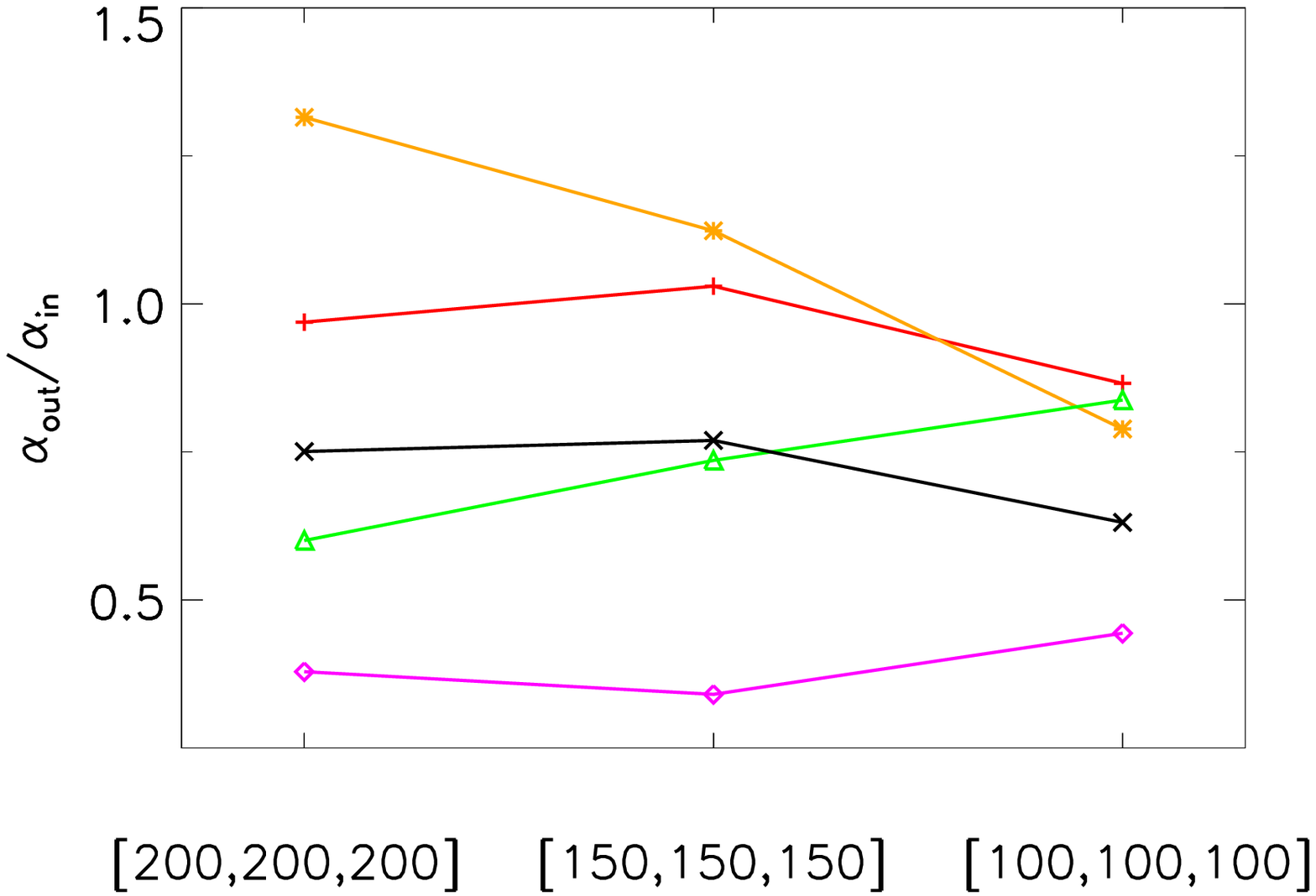}}
\caption{The ratio between the output virial parameter ($\alpha_{out}$) derived from the six algorithms and the input virial parameter ($\alpha_{in}$) for clumps with different size (a), SNR (b), and crowdedness (c).}
\label{fig:vir_para}
\end{figure}

The clumps which are in gravitationally unstable will collapse and are expected to evolve into protostars. The virial parameter $\alpha_{vir}$, which is defined as $\alpha_{vir} = 5\sigma^2 R_c/(GM_c) \sim E_{kin}/E_{g}$ \citep{1992ApJ...395..140B}, is a crucial parameter to understand the dynamics of a clump. Here, $E_{kin}$ and $E_{g}$ are the kinematic and gravitational energy of the clump, respectively. $M_c$ and $R_c$ indicate the mass and radius of the clump, $G$ is the gravitational constant, and $\sigma$ is the velocity dispersion. In the absence of external pressure or magnetic fields for an isothermal sphere, the clump will collapse if $\alpha_{vir} < 1$ \citep{2014prpl.conf..149T}. When the region is under external pressure, this pressure will work towards compressing the clump, and it could collapse even if $\alpha_{vir} > 1$. Several surveys revealed a relationship between $\alpha_{vir}$ and the mass of the clumps, indicating that the massive clumps tend to be more gravitationally unstable \citep{2014MNRAS.443.1555U,2018MNRAS.473.1059U,2018MNRAS.477.2220T}. However, the $\alpha_{vir}$ derived from the six algorithms can be influenced by the bias of the returned parameters. Here, we assume that the mass of the clump is proportional to the total flux of the clump ($M_c \sim flux$). The ratio between the output virial parameter ($\alpha_{out}$) derived from the six algorithms and the input virial parameter ($\alpha_{in}$) is displayed in Figure \ref{fig:vir_para}.

As shown in Figure \ref{fig:vir_para}, the six algorithms almost return a virial parameter close to the input virial value. The output virial parameter shows no trend with the size, SNR, and crowdedness of the clumps. ClumpFind1994 and Dendrograms return more accurate virial parameter than the other algorithms.

\subsection{The Performance of the Algorithms in Identifying Clumps in the Rosette Molecular Cloud}
\label{discussion2}

\begin{figure} [!htb]
\centering
\subfloat[]{\label{fig:12rosette}
\includegraphics[width=0.3\textwidth]{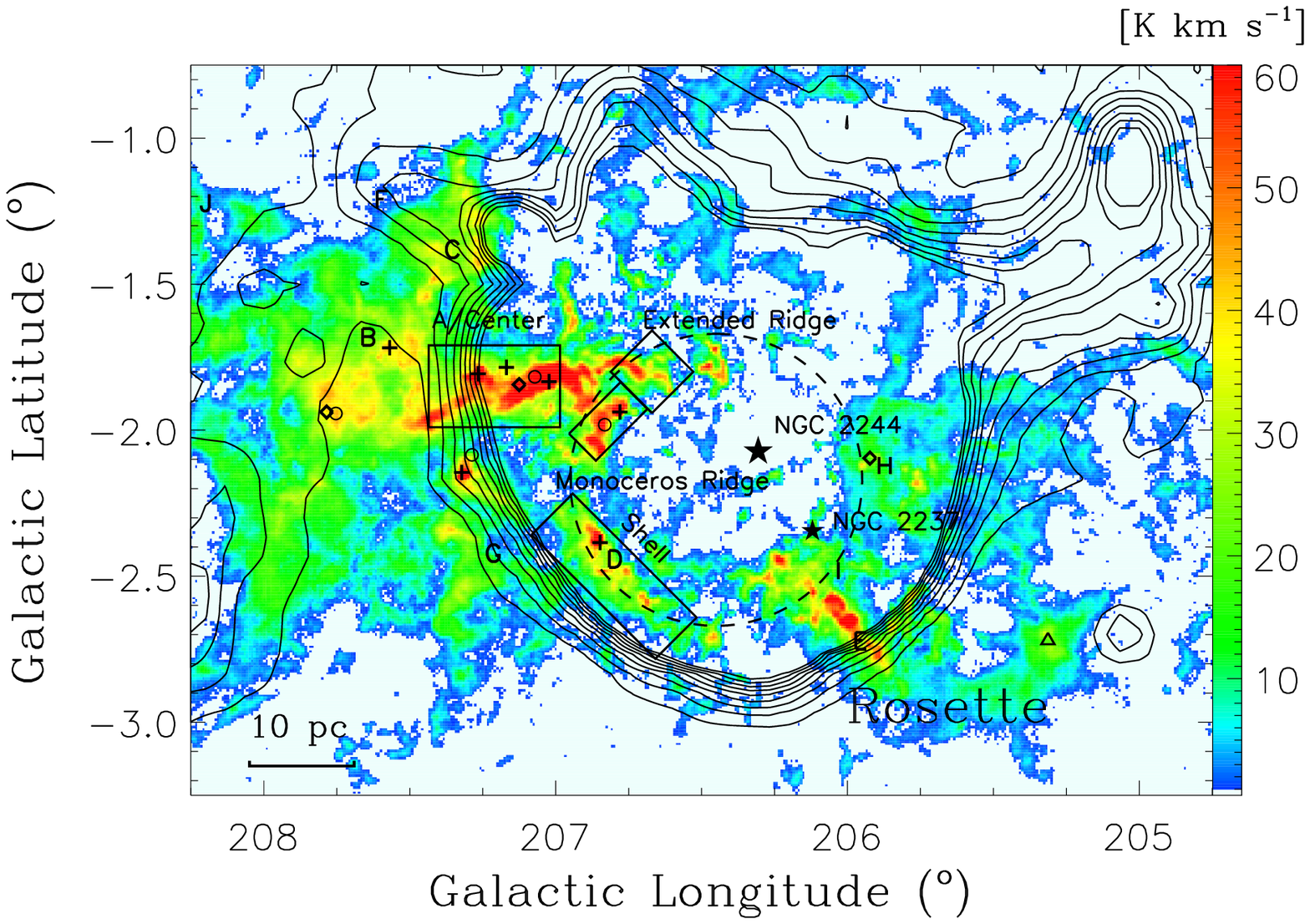}}
\subfloat[]{\label{fig:13rosette}
\includegraphics[width=0.3\textwidth]{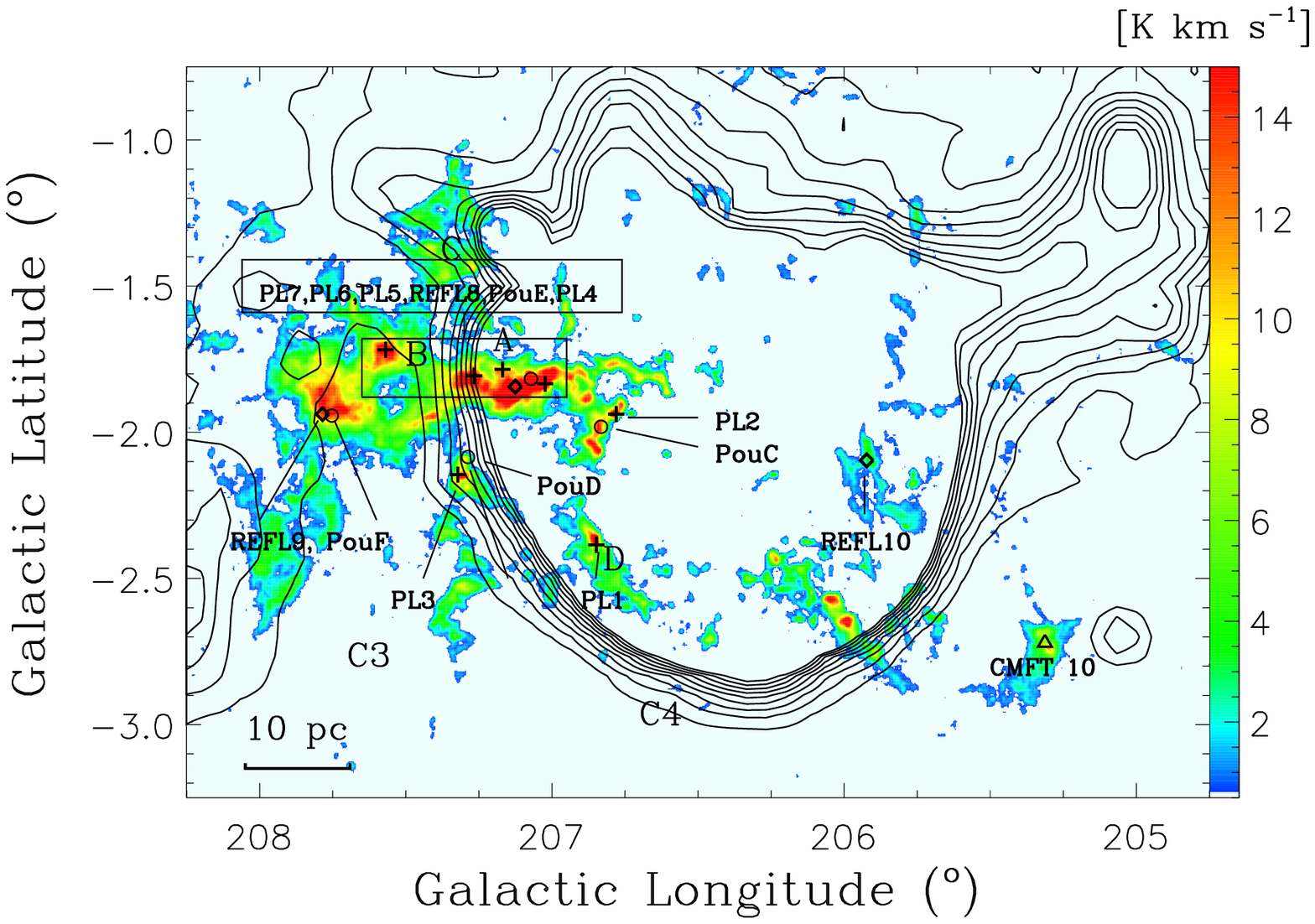}}
\subfloat[]{\label{fig:18rosette}
\includegraphics[width=0.3\textwidth]{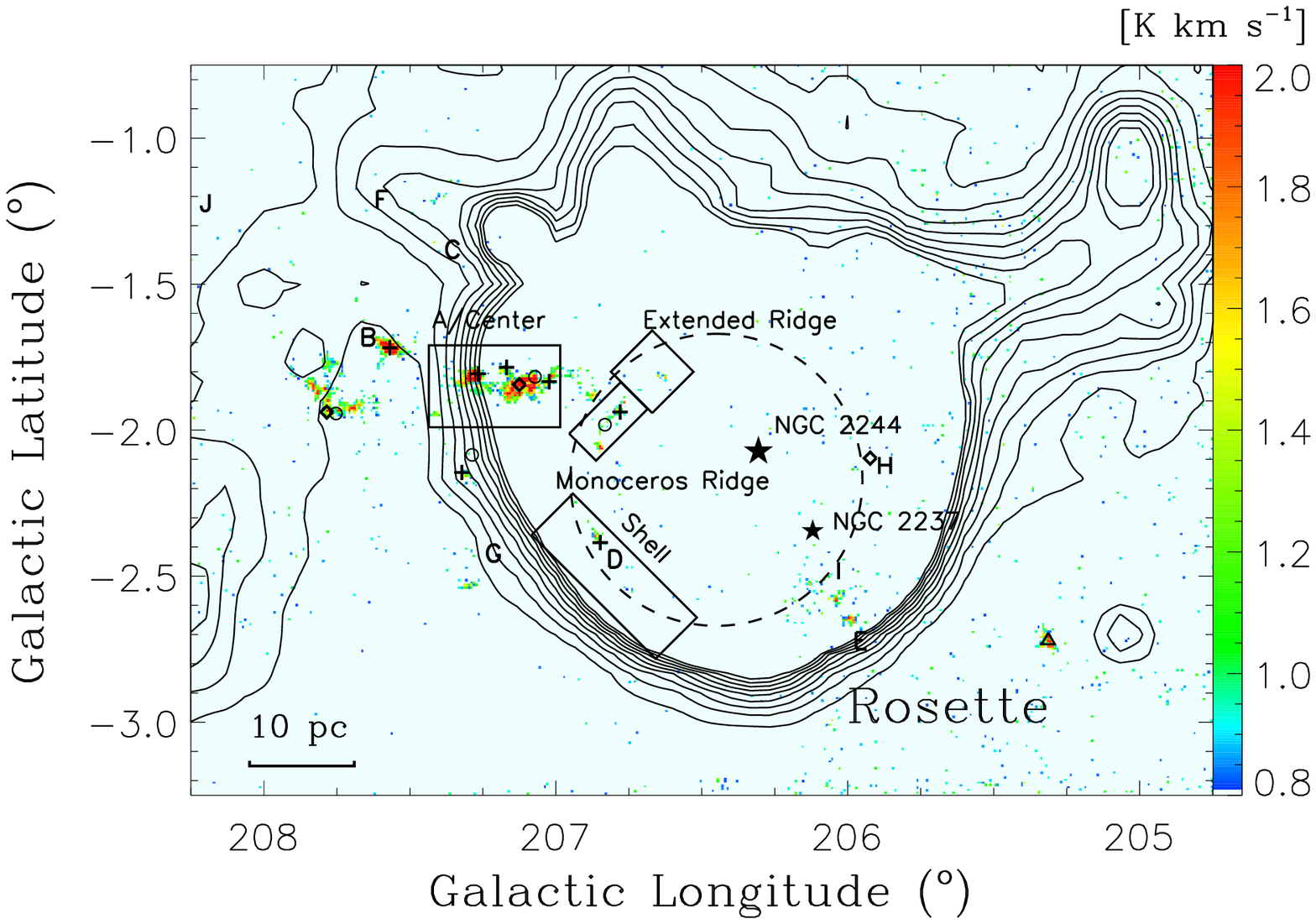}}
\caption{Maps of $^{12}$CO emission intensity integrated from -2 km s$^{-1}$ to 30 km s$^{-1}$ (a), $^{13}$CO emission intensity integrated from 3 km s$^{-1}$ to 26 km s$^{-1}$ (b), and C$^{18}$O emission intensity integrated from 3 km s$^{-1}$ to 19 km s$^{-1}$ (c). For details, see \citet{2018ApJS..238...10L}.}
\label{fig:3rosette}
\end{figure}

\begin{figure} [!htb]
\centering
\subfloat[]{\label{fig:total_rosette} 
\includegraphics[width=0.4\textwidth]{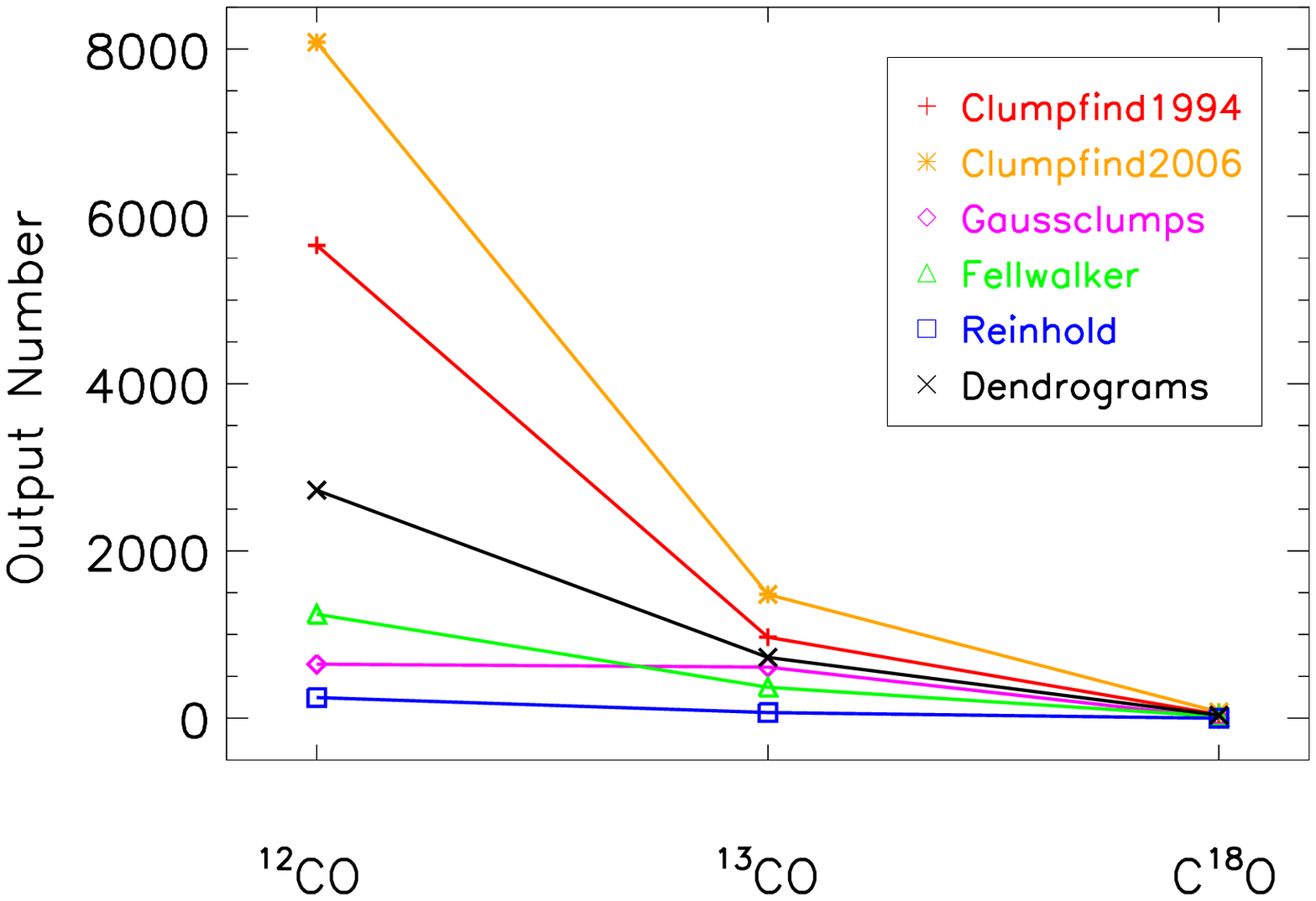}}
\subfloat[]{\label{fig:time_rosette}
\includegraphics[width=0.4\textwidth]{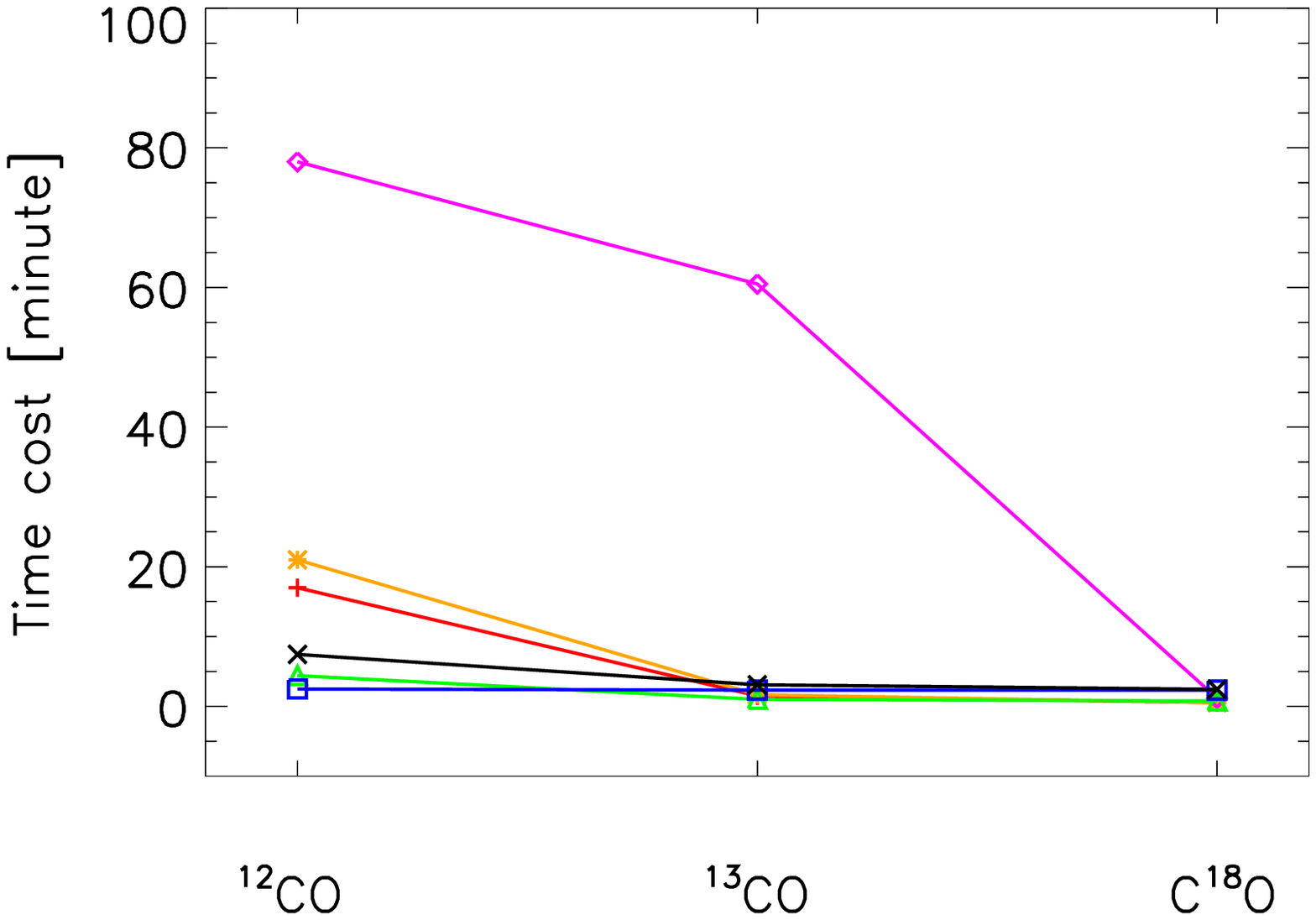}}\\
\caption{Left: the output number for each algorithm in identifying clumps of $^{12}$CO, $^{13}$CO, and C$^{18}$O emission. Right: the CPU consuming time of the six algorithms.}
\label{fig:rosette}
\end{figure}

\begin{figure} [!htb]
\centering
\subfloat[]{\label{fig:size_rosette}
\includegraphics[width=0.4\textwidth]{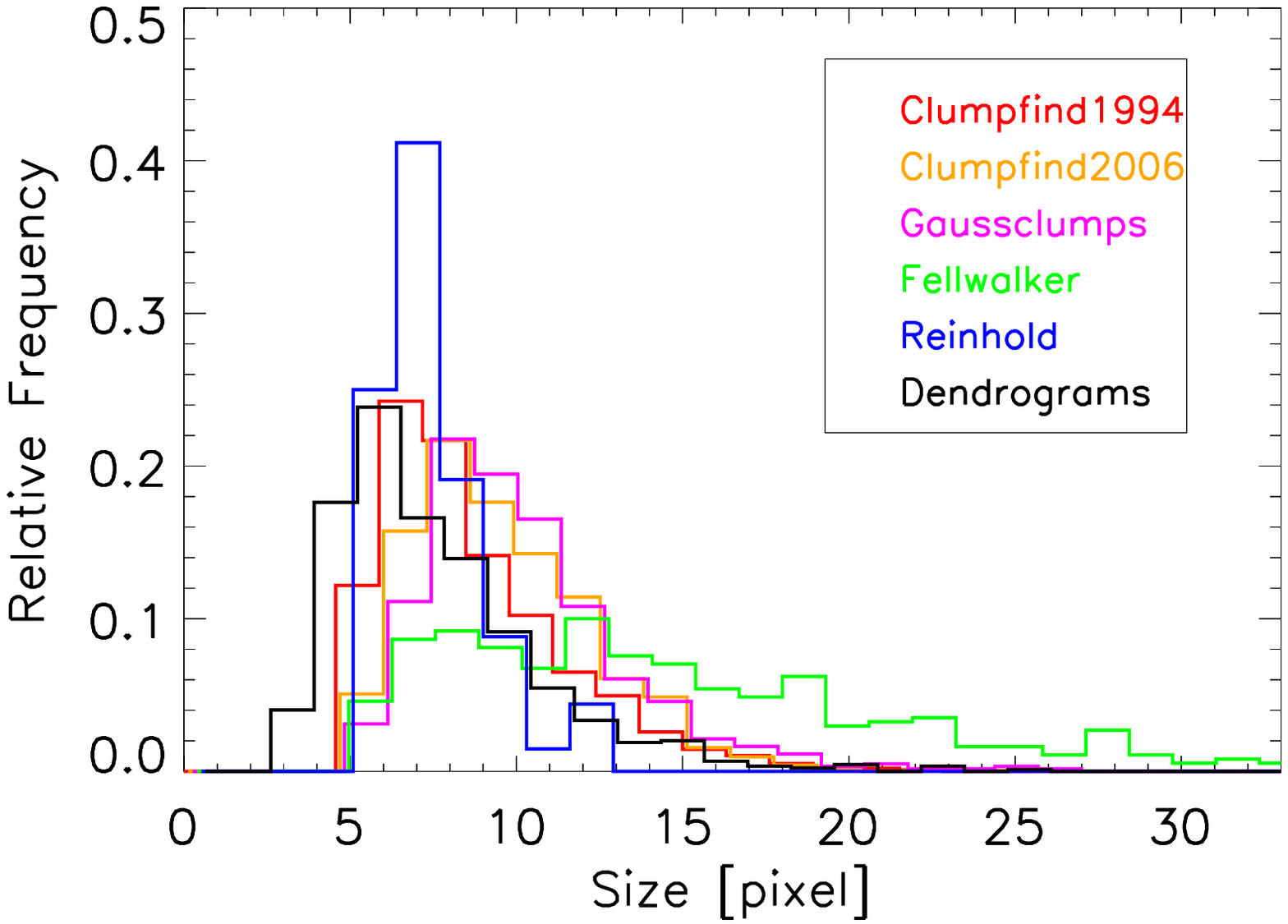}}
\subfloat[]{\label{fig:deltv_rosette}
\includegraphics[width=0.4\textwidth]{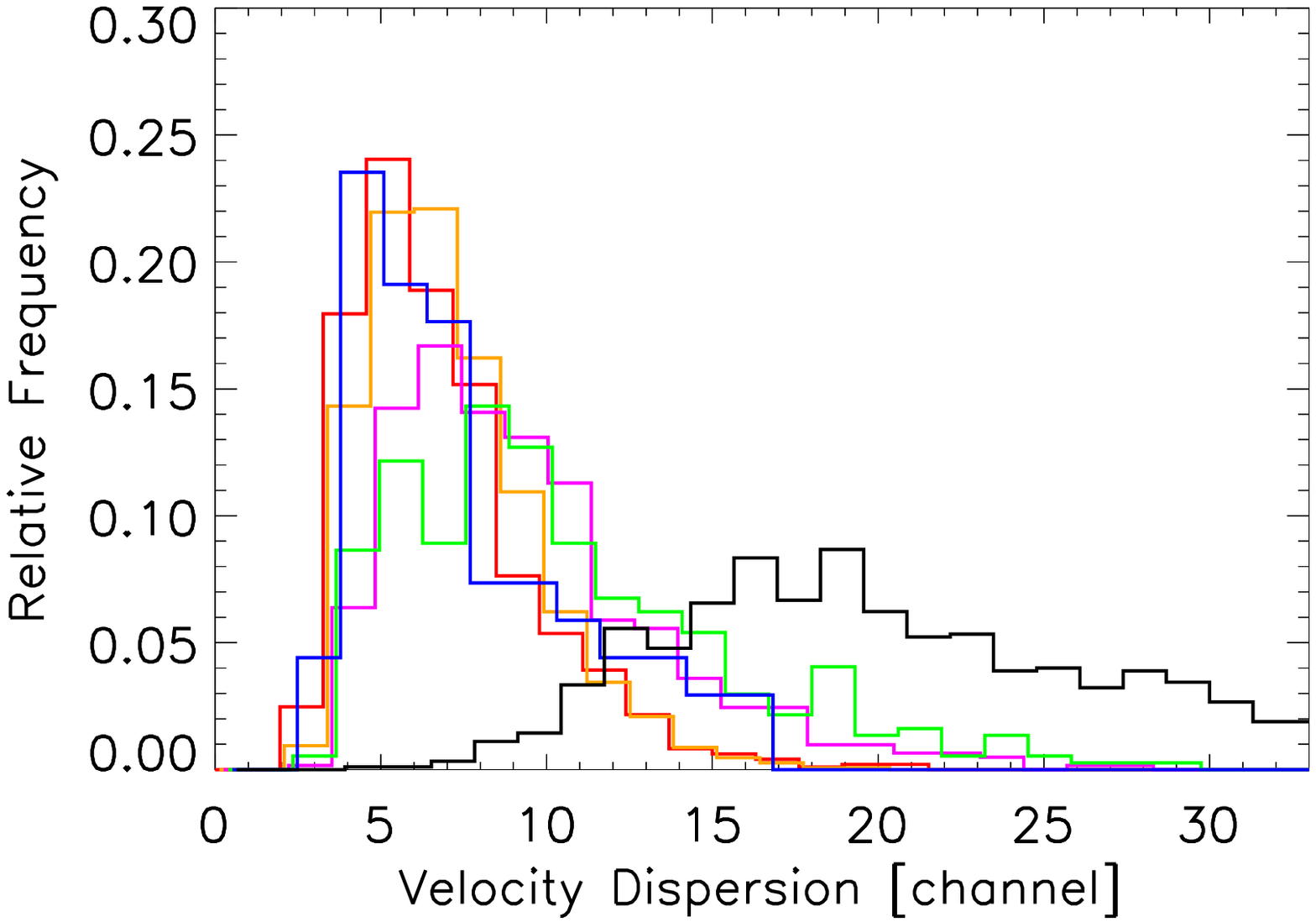}}\\
\subfloat[]{\label{fig:peak_rosette}
\includegraphics[width=0.4\textwidth]{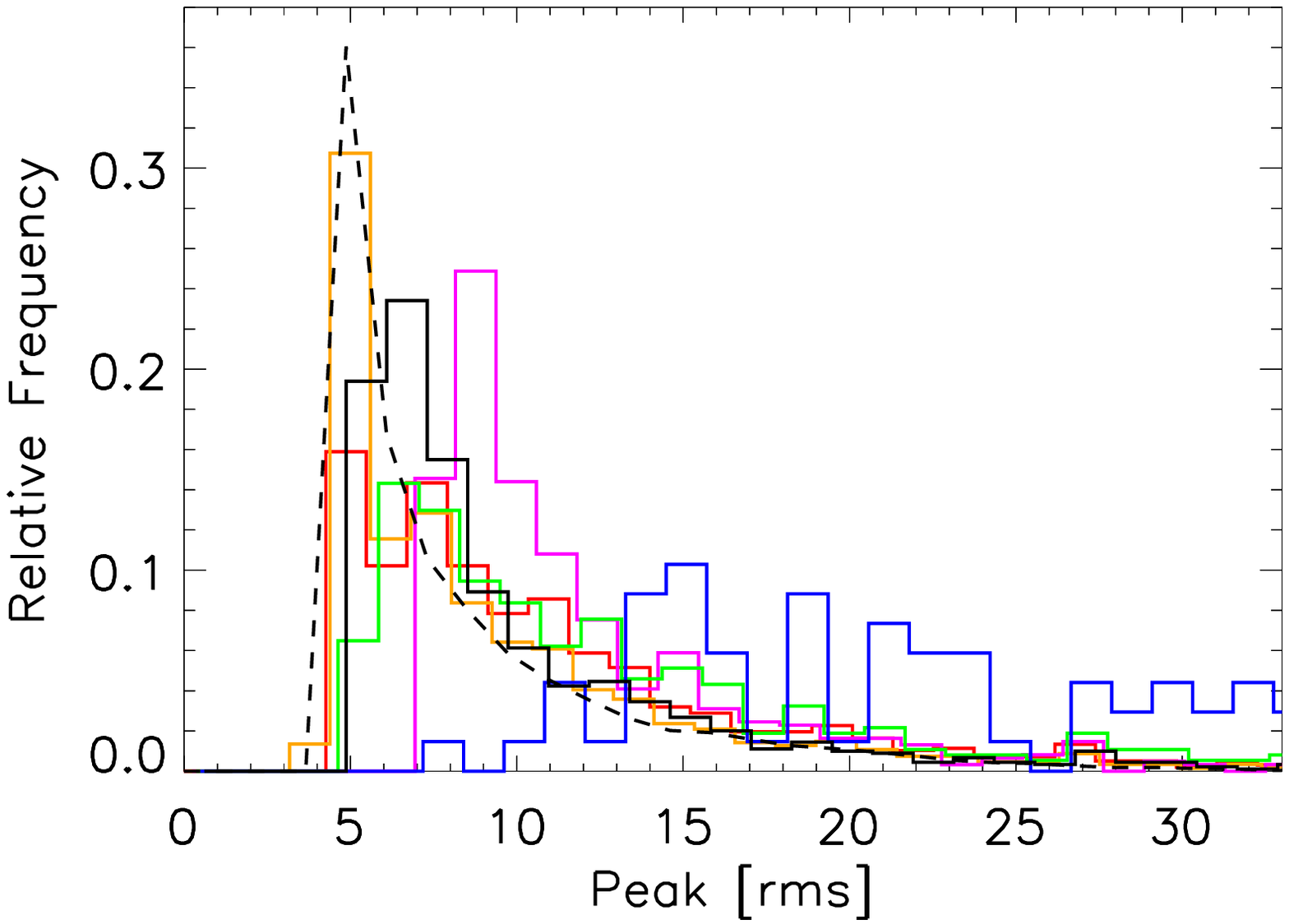}}
\subfloat[]{\label{fig:sum_rosette}
\includegraphics[width=0.4\textwidth]{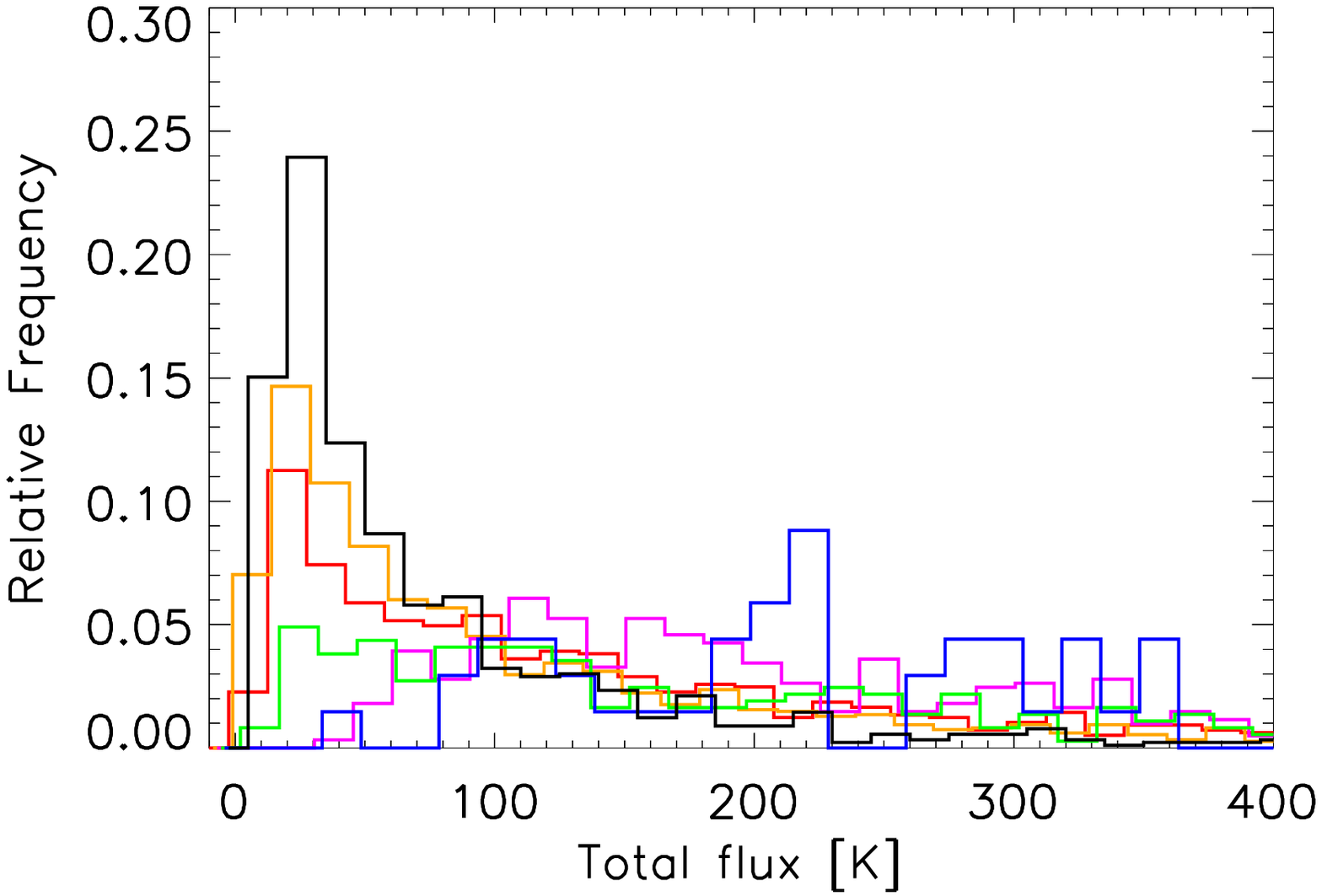}}
\caption{Distributions of the size (a), velocity dispersion (b), peak brightness (c), and total flux (d) of the $^{13}$CO clumps in the RMC derived from the six algorithms. The dashed line indicates the probability distribution of the pixel peak brightness of the RMC.}
\label{fig:hist_rosette}
\end{figure}

Using the PMO-13.7m millimeter telescope at Delingha in China, \citet{2018ApJS..238...10L} have conducted a large-scale simultaneous survey of $^{12}$CO, $^{13}$CO, and C$^{18}$O J=1-0 emission toward the Rosette molecular cloud (RMC) region with a sky coverage of $3.5 \times 2.5$ square degrees (Figure \ref{fig:3rosette}). The spatial pixel of the FITS data cube has a size of 30$\arcsec$ $\times$ 30$\arcsec$ and the effective spectral resolution is 61.0 kHz, corresponding to a velocity resolution of 0.16 km s$^{-1}$ at the 115 GHz frequency of the $^{12}$CO $J=1-0$ line. The sensitivity of the observation is estimated to be around 0.5 K for the $^{12}$CO $J=1-0$ emission and around 0.3 K for the $^{13}$CO and C$^{18}$O $J=1-0$ emission. We apply the six algorithms in identifying clumps in the RMC, and the results are presented in Figures \ref{fig:rosette}-\ref{fig:hist_rosette}. 

Figure \ref{fig:total_rosette} presents the output number for each algorithm in identifying clumps of $^{12}$CO, $^{13}$CO, and C$^{18}$O emission. It shows that ClumpFind2006 and ClumpFind1994 get the most clumps in the RMC. As shown in the above simulated tests (Section \ref{result}), ClumpFind2006 and ClumpFind1994 always get much more repeated and erroneous identifications than the number of input simulated clumps, producing more output clumps than other algorithms. Reinhold gets the least clumps in the RMC due to that it can only find clumps when the peak brightness is higher than 12 times the noise level (see Section \ref{result} and Figure \ref{fig:peak_rosette}). Gaussclumps gets relatively fewer clumps of $^{12}$CO, for which one possible reason is that $^{12}$CO emission is usually optically thick and the velocity profiles of $^{12}$CO clumps do not obey Gaussian profile well. The iterative fitting process of Gaussclumps is terminated when more than designated consecutive clumps cannot be fitted with Gaussian profile successfully. For Dendrograms, it rarely finds false clumps in the simulated data (see Section \ref{result}). However, we find that more than 100 false clumps are identified by Dendrograms, but these clumps are distributed at the edge of observational $^{12}$CO, $^{13}$CO, and C$^{18}$O data arrays. The numbers shown in Figure \ref{fig:total_rosette} and Table \ref{tab1} do not include the false clumps located at the edges of th data arrays. Figure \ref{fig:time_rosette} presents the computer CPU consuming time for the six algorithms in identifying $^{12}$CO, $^{13}$CO, and C$^{18}$O clumps. Reinhold takes the least time and the least clumps are identified. Among the six algorithms, Fellwalker is the most efficient algorithm in terms of the number of identified clumps and CPU time. Due to the iterative Gaussian fitting process, the time cost by Gaussclumps is much more than the other algorithms. The dominant frequency of the computer CPU in our testing is 2.2 GHz. The memory size of the computer is 16 GB and the memory speed is 1600 MHz.

\begin{table}[!htb]
\bc
\caption[]{Cross-matching of $^{12}$CO, $^{13}$CO, and C$^{18}$O clumps}\label{tab1}
\setlength{\tabcolsep}{1pt}
\scriptsize
 \begin{tabular}{lcccccccccccc}
  \hline\noalign{\smallskip}
    \hline\noalign{\smallskip}
Algorithm && $^{12}$CO && $^{13}$CO && C$^{18}$O &&&  $^{13}$CO/$^{12}$CO &&& C$^{18}$O/$^{13}$CO \\
  \hline\noalign{\smallskip}
ClumpFind1994           && 5653       &&  969       &&   44   &&& 76$\%$                                   &&& 55$\%$                         \\
ClumpFind2006           && 8080       &&  1480     &&   75   &&& 74$\%$                                   &&& 37$\%$                         \\
Gaussclumps              && 646         &&   611       &&   4     &&& 67$\%$                                   &&& 100$\%$                       \\
Fellwalker                   && 1243        &&  370       &&   22   &&& 81$\%$                                   &&& 91$\%$                          \\
Reinhold                     && 247         &&   68         &&   0     &&& 15$\%$                                   &&& -                                     \\
Dendrograms             && 2726        &&  727       &&   36   &&& 92$\%$                                   &&& 0$\%$                                    \\
  \noalign{\smallskip}\hline
\end{tabular}
\ec
\tablecomments{0.6\textwidth}{Columns 2-4 give the output number of the clumps. Columns 5-6 give the percentage of the $^{13}$CO clumps that coindcide with $^{12}$CO clumps and the percentage of the C$^{18}$O clumps that coindcide with $^{13}$CO clumps, respectively.}
\end{table}

Table \ref{tab1} presents the cross-matching results between $^{12}$CO, $^{13}$CO, and C$^{18}$O clumps. Due to the difference in optical depth between the $^{12}$CO, $^{13}$CO, and C$^{18}$O emission, it is usually expected that $^{13}$CO clumps have good association with $^{12}$CO clumps and C$^{18}$O clumps in the same way have good association with $^{13}$CO clumps. From Table \ref{tab1} it can be seen that the $^{13}$CO and C$^{18}$O clumps from Gaussclumps and Fellwalker exhibit the best association. However, Gaussclumps identifies only four C$^{18}$O clumps, which is much fewer than the number identified by eyes. For ClumpFind1994, ClumpFind2006, Reinhold, and Dendrograms, the association between the $^{13}$CO and C$^{18}$O clumps is relatively low, which implies that their identifications deviate somewhat from the actual situation.

Figure \ref{fig:hist_rosette} shows the distributions of the size, velocity dispersion, peak brightness, and total flux of $^{13}$CO clumps, respectively. It can be seen that the clump size identified by Fellwalker is larger than that from the other algorithms. The most likely reason is that Fellwalker returns the larger and more accurate clump size than the other algorithms (see Section \ref{result}). The clump velocity dispersion identified by Dendrograms is significantly larger than that from the other algorithms. As shown in Figure \ref{fig:peak_rosette}, most of the clumps identified by Gaussclumps and Reinhold have higher peak brightness than the clumps from the other algorithms, which is consistent with the test results in Section \ref{test2} that only clumps with high brightness can be identified by Gaussclumps and Reinhold. The distribution of the total flux of $^{13}$CO clumps is presented in Figure \ref{fig:sum_rosette}. Due to that Fellwalker returns a larger and more accurate clump total flux than the other algorithms, Gaussclumps and Reinhold tend to miss the clumps with low brightness, it can be seen that the total flux extracted by Reinhold, Gaussclumps, and Fellwalker are higher than that by the other algorithms.

\section{Summary}
\label{summary}

Using simulated clumps, we have tested the performance of the GaussClumps, ClumpFind1994, ClumpFind2006, Fellwalker, Reinhold, and Dendrograms algorithms in identifying clumps. We focus on the performance of each algorithm in terms of completeness and parameter extraction. 

We generated the simulated clumps in three-dimensional arrays with background noise. The brightness profiles of the clumps are of the form $T(s,v) = \frac{N_0}{\sqrt{2\pi}\sigma} \times \frac{1}{1+(s/r_0)^{1.0}} \times exp(\frac{-(v-v_0)^2}{2\sigma^2})$. The simulated clumps are designed to vary in size, brightness, and crowdedness in order to investigate the performance of the six algorithms in these aspects. For the six algorithms, the minimum FWHM and $\Delta$V of the identifying clumps are set to be 2 pixels and 2 channels, respectively. The minimum peak value parameter is set to be 5 times the one sigma noise level and the number of voxels of an output clump is required to be above 16. We summarize our results as follows,

1. In the aspect of detection completeness, Fellwalker, Dendrograms, and Gaussclumps are the first, second, and third best algorithms, respectively. The numbers of correct identifications of the six algorithms gradually increase as the size and SNR of the simulated clumps increase and they decrease as the crowdedness increases. The repetitive and erroneous rates of ClumpFind increase as the clump size and SNR increase. Reinhold is only suitable for searching for clumps with peak brightness (SNR) higher than 10. The general performances of the six algorithms are summarized in Table \ref{tab2}.

\newcommand{\tabincell}[2]{\begin{tabular}{@{}#1@{}}#2\end{tabular}}
\begin{table}[!htb]
\bc
\caption[]{General Performance of the Algorithms in Completeness}\label{tab2}
\setlength{\tabcolsep}{1pt}
\scriptsize
 \begin{tabular}{lcccccccc}
  \hline\noalign{\smallskip}
    \hline\noalign{\smallskip}
Algorithm && Correct Rate   && Repetitive Rate   &&  Erroneous Rate \\
  \hline\noalign{\smallskip}
ClumpFind1994 &&   high (100$\%$) && very high ($<10000\%$) && very high ($<15000\%$)\\
ClumpFind2006 &&   high (100$\%$) && very high ($<12000\%$) && very high ($<20000\%$)\\
Gaussclumps &&   intermediate ($10\%-100\%$) && low ($<10\%$)  && intermediate  ($<70\%$) \\
Fellwalker &&   high ($90\%-100\%$) && very low (0$\%$) && very low (0$\%$)\\
Reinhold &&   low ($<10\%$ when clump peak brightness (SNR) lower than 15) && very low (0$\%$) && very low (0$\%$)\\
Dendrograms &&   high ($100\%$) && high  ($<1000\%$) && high  ($<2000\%$)\\
  \noalign{\smallskip}\hline
\end{tabular}
\ec
\end{table}

2. In the aspect of the accuracy of retrieved parameters, the average errors in clump parameters gradually increase as the clump size, SNR, and crowdedness increase. The average errors of the algorithms in extracting parameters of the clumps in Section \ref{test1}-\ref{test3} are presented in Table \ref{tab3}. As Table \ref{tab3} shows, the algorithms best performing in extracting parameters are Dendrograms in retrieving clump position ( E($|\Delta$X$|$)=$0.4$ pixels ) and peak brightness ( E($\Delta$I)=$1.3$ RMS), Fellwalker in size ( E($\Delta$S)=$-7\%$ ) and total flux ( E($\Delta$flux) =$-19\%$), and ClumpFind1994 in velocity dispersion (E($\Delta$V)=$-38\%$). All in all, Fellwalker, Dendrograms, and Gaussclumps exhibit better performance in extracting clump parameters than the other algorithms. Except for Fellwalker, the other algorithms exhibit significant deviation in extracting the total flux of clumps.

\begin{table}[!htb]
\bc
\caption[]{Average Errors of the Algorithms in Extracting Parameters}\label{tab3}
\setlength{\tabcolsep}{1pt}
\scriptsize
 \begin{tabular}{lcccccccc}
  \hline\noalign{\smallskip}
    \hline\noalign{\smallskip}
Algorithm & ~~~E($|\Delta$X$|$)   && ~~~E($\Delta$S) && ~~~E($\Delta$V)   &&  ~~~E($\Delta$I)  &~~~E($\Delta$flux) \\
                &   ~~~(pixels)                &&                        &&                          &&   ~~~(rms)            &                         \\
  \hline\noalign{\smallskip}
ClumpFind1994 &   $1.1$ && $-38\%$  &&$-38\%$ && $1.3$ & $-70\%$\\
ClumpFind2006 &   $1.1$ && $-57\%$  &&$-40\%$ && $1.3$ & $-83\%$\\
Gaussclumps &      $1.1$ && $-65\%$  &&$-58\%$  && $1.5$ & $-78\%$\\
Fellwalker &            $0.9$ &&  $-7\%$   &&$-43\%$ &&  $1.8$ & $-19\%$\\
Reinhold &              $1.8$ &&  $-88\%$ &&$-53\%$ &&  $2.9$ & $-93\%$\\
Dendrograms &      $0.4$ &&  $-25\%$ &&$-42\%$ &&  $1.3$ & $-65\%$\\
  \noalign{\smallskip}\hline
\end{tabular}
\ec
\tablecomments{0.6\textwidth}{The average errors of the algorithms in extracting parameters are derived from the (1000+1000+300) clumps in Section \ref{test1}-\ref{test3}. Column 2 and 5 represent the average deviation of the position (E($|\Delta$X$|$)) and peak brightness (E($\Delta$I)). Column 3, 4, and 6 give the relative errors of the size (E($\Delta$S)), velocity dispersion (E($\Delta$V)), and total flux (E($\Delta$flux)).}
\end{table}

3.  The ratios between the output virial parameter ($\alpha_{out}$) derived from the six algorithms and the input virial parameter ($\alpha_{in}$) show no trend with the size, SNR, and crowdedness of the clumps. For the simulated clumps, the six algorithms almost return virial parameters similar to the input virial parameters ($\alpha_{out}/\alpha_{in} = 0.5 - 1.5$). 

4. When applying the six algorithms to clump identification for the RMC, Gaussclumps, ClumpFind1994, ClumpFind2006, Fellwalker, and Reinhold exhibit performance that is consistent with the results from the simulated test. Dendrograms finds more than 100 false clumps at the edge of observational data arrays.

\normalem
\begin{acknowledgements}
We would like to thank the MWISP members for the helpful discussions. We thank the anonymous referee for valuable comments and suggestions that helped to improve this paper. This work is supported by the National Key R\&D Program of China (NO. 2017YFA0402701). C. Li acknowledges supports by NSFC grants 11503086 and 11503087. 
\end{acknowledgements}

\clearpage
\appendix
\section{Algorithms Configuration Parameters}          
\label{appendix}

\begin{table}[h]
\bc
\caption[]{GaussClumps Parameters}
\setlength{\tabcolsep}{1pt}
\scriptsize
 \begin{tabular}{lcccccccccccc}
  \hline\noalign{\smallskip}
   GAUSSCLUMPS.EXTRACOLS=0\\
   GAUSSCLUMPS.FWHMBEAM=2\\
   GAUSSCLUMPS.MAXBAD=0.05\\
   GAUSSCLUMPS.MAXCLUMPS=2147483647\\
   GAUSSCLUMPS.MAXNF=100\\
   GAUSSCLUMPS.MAXSKIP=10\\
   GAUSSCLUMPS.MAXWF=1.1\\
   GAUSSCLUMPS.MINPIX=3\\
   GAUSSCLUMPS.MINWF=0.8\\
   GAUSSCLUMPS.MODELLIM=3\\
   GAUSSCLUMPS.NPAD=10\\
   GAUSSCLUMPS.NPEAK=9\\
   GAUSSCLUMPS.NSIGMA=3\\
   GAUSSCLUMPS.NWF=10\\
   GAUSSCLUMPS.S0=1\\
   GAUSSCLUMPS.SA=1\\
   GAUSSCLUMPS.SB=0.1\\
   GAUSSCLUMPS.SC=1\\
   GAUSSCLUMPS.THRESH=5\\
   GAUSSCLUMPS.VELORES=2\\
   GAUSSCLUMPS.WMIN=0.05\\
   GAUSSCLUMPS.WWIDTH=2\\
  \noalign{\smallskip}\hline
\end{tabular}
\ec
\end{table}

\begin{table}[h]
\bc
\caption[]{ClumpFind1994 Parameters}
\setlength{\tabcolsep}{1pt}
\scriptsize
 \begin{tabular}{lcccccccccccc}
  \hline\noalign{\smallskip}
   CLUMPFIND.ALLOWEDGE=0\\
   CLUMPFIND.DELTAT=2*RMS\\
   CLUMPFIND.FWHMBEAM=2\\
   CLUMPFIND.IDLALG=0\\
   CLUMPFIND.MAXBAD=0.05\\
   CLUMPFIND.MINPIX=16\\
   CLUMPFIND.NAXIS=3\\
   CLUMPFIND.Noise=2*RMS\\
   CLUMPFIND.TLOW=3*RMS\\
   CLUMPFIND.VELORES=2\\
  \noalign{\smallskip}\hline
\end{tabular}
\ec
\end{table}

\begin{table}
\bc
\caption[]{ClumpFind2006 Parameters}
\setlength{\tabcolsep}{1pt}
\scriptsize
 \begin{tabular}{lcccccccccccc}
  \hline\noalign{\smallskip}
   CLUMPFIND.ALLOWEDGE=0\\
   CLUMPFIND.DELTAT=2*RMS\\
   CLUMPFIND.FWHMBEAM=2\\
   CLUMPFIND.IDLALG=1\\
   CLUMPFIND.MAXBAD=0.05\\
   CLUMPFIND.MINPIX=16\\
   CLUMPFIND.NAXIS=3\\
   CLUMPFIND.Noise=2*RMS\\
   CLUMPFIND.TLOW=3*RMS\\
   CLUMPFIND.VELORES=2\\
  \noalign{\smallskip}\hline
\end{tabular}
\ec
\end{table}

\begin{table}
\bc
\scriptsize
\caption[]{ Fellwalker Parameters}
\setlength{\tabcolsep}{1pt}
 \begin{tabular}{lcccccccccccc}
  \hline\noalign{\smallskip}
   FELLWALKER.ALLOWEDGE=1\\
   FELLWALKER.CLEANITER=1\\
   FELLWALKER.FLATSLOPE=1*RMS\\
   FELLWALKER.FWHMBEAM=2\\
   FELLWALKER.MAXBAD=0.05\\
   FELLWALKER.MAXJUMP=4\\
   FELLWALKER.MINDIP=3*RMS\\
   FELLWALKER.MINHEIGHT=5*RMS\\
   FELLWALKER.MINPIX=16\\
   FELLWALKER.NOISE=3*RMS\\
   FELLWALKER.VELORES=2\\
  \noalign{\smallskip}\hline
\end{tabular}
\ec
\end{table}

\begin{table}
\bc
\caption[]{Reinhold Parameters}
\setlength{\tabcolsep}{1pt}
\scriptsize
 \begin{tabular}{lcccccccccccc}
  \hline\noalign{\smallskip}
   REINHOLD.CAITERATIONS=1\\
   REINHOLD.CATHRESH=26\\
   REINHOLD.FIXCLUMPSITERATIONS=1\\
   REINHOLD.FLATSLOPE=1*RMS\\
   REINHOLD.FWHMBEAM=2\\
   REINHOLD.MINLEN=4\\
   REINHOLD.MINPIX=16\\
   REINHOLD.NOISE=3*RMS\\
   REINHOLD.THRESH=5*RMS\\
   REINHOLD.VELORES=2\\
  \noalign{\smallskip}\hline
\end{tabular}
\ec
\end{table}

\begin{table}
\bc
\caption[]{Dendrograms Parameters}
\setlength{\tabcolsep}{10pt}
\scriptsize
 \begin{tabular}{lcccccccccccc}
  \hline\noalign{\smallskip}
   MIN\_VALUE=3*RMS\\
   MIN\_DELTA=2*RMS\\
   MIN\_NPIX=16\\
  \noalign{\smallskip}\hline
\end{tabular}
\ec
\end{table}

\clearpage  
\bibliographystyle{raa}
\bibliography{refs}

\end{document}